\documentclass[twocolumn]{autart}
\usepackage{amsmath,amssymb,amsfonts}
\usepackage{mathrsfs}

\usepackage{algorithmic}
\usepackage{graphicx}
\usepackage{booktabs}
\usepackage{algorithm}
\usepackage[caption=false,font=footnotesize]{subfig}
\begin{document}

\begin{frontmatter}
\runtitle{Tube-Based Robust Data-Driven Predictive Control for LPV Systems}
\title{Tube-Based Robust Data-Driven Predictive Control for Linear Parameter-Varying Systems With Noisy Data\thanksref{footnoteinfo}} \thanks[footnoteinfo]{This paper was not presented at any IFAC meeting. Corresponding author: Chi Wang.}
\author[Imperial]{Chi Wang}\ead{chi.wang23@imperial.ac.uk},
\author[Imperial,Florence]{David Angeli}\ead{d.angeli@imperial.ac.uk, david.angeli@unifi.it}
\address[Imperial]{Department of Electrical and Electronic Engineering, Imperial College London, London SW7 2AZ, U.K.}\address[Florence]{Department of Information Engineering, University of Florence, Italy}


\begin{abstract}
This paper proposes a tube-based robust data-driven predictive control scheme
for unknown discrete-time linear parameter-varying (LPV) systems using offline
noisy input-scheduling-state data. We establish a multi-trajectory simplified
LPV fundamental lemma yielding a collective Hankel representation of
finite-horizon trajectories. Constraining the Hankel coefficients to a scaled
simplex uniformly bounds the noise-induced predictor mismatch but may exclude
relevant constrained lifted trajectories. A convex-hull coverage condition for
admissible receding-horizon continuations restores the required
representability. LPV data-consistency and robust positively invariant sets
provide tube-tightening ingredients computed directly from noisy data. The
online problem is a quadratic program, for which recursive feasibility,
closed-loop constraint satisfaction, and practical input-to-state stability
are established. Numerical results illustrate the effectiveness of the proposed scheme.
\end{abstract}

\end{frontmatter}

\small

\section{Introduction}
\label{sec:introduction}
Data-driven control has become a powerful paradigm when accurate
first-principles models are unavailable or prohibitively costly to obtain.
Substantial progress has been made in direct data-driven controller design for
unconstrained problems, including state-feedback
stabilization~\cite{bisoffi2021trade,DePersisTesi2020TAC,verhoek2025direct,WangAngeli2026CDCsub},
data-driven $\mathcal H_2$ control~\cite{vanWaarde2020IFAC}, and data-driven
linear quadratic regulation~\cite{Dorfler2023TAC}. By contrast, constrained
control remains more challenging, despite its importance in safety-critical
applications. Model predictive control
(MPC)~\cite{Mayne2005,RawlingsMayneDiehl2017}, with its ability to handle
constraints while optimizing closed-loop performance online, is therefore a
natural framework for constrained data-driven control.

Much of the existing literature on MPC relies on an explicit model that is
first obtained or identified and then used for predictive
control~\cite{Abbas2016TAC,Gonzalez2020IJRNC,RossiterKouvaritakis2001,Terzi2019Automatica}.
A conceptually different route is offered by direct data-driven predictive
control rooted in behavioral system theory. In the linear time-invariant (LTI)
setting, Willems' fundamental lemma~\cite{Willems2005} states that, if the
input of a noise-free measured trajectory is persistently exciting of a
sufficiently high order, then all trajectories of a prescribed length lie in
the column space of the associated data Hankel
matrix~\cite{berberich2020trajectory}. Building on this result, data-enabled
predictive control (DeePC) was proposed as a behavioral data-driven alternative
to model-based MPC~\cite{coulson2019data}.

In its classical form, the exact behavioral parametrization underlying
Willems' fundamental lemma applies to noise-free LTI systems. Extensions of
the fundamental lemma and related behavioral representations have been
developed for richer classes, including linear time-varying
systems~\cite{Nortmann2023TAC}, feedback-linearizable
systems~\cite{Alsalti2023TAC}, and linear parameter-varying (LPV)
systems~\cite{Verhoek2026TAC_behavioral}. Direct data-driven predictive
control schemes have also been proposed for such
classes~\cite{Alsalti2023IFAC,Hu2024CSL,Verhoek2026TAC}. These developments,
however, remain largely tied to noise-free data or do not provide robust
closed-loop guarantees for controllers built directly from noisy data.

In the LTI setting, robust data-driven MPC formulations with closed-loop
guarantees include approaches based on terminal equality
constraints~\cite{Berberich2020,Berberich2021TAC}, min--max
formulations~\cite{xie2026}, quadratic-matrix-inequality
formulations~\cite{Hu2025TAC}, and tube-based
constructions~\cite{WangAngeli2026TRDDPC}. Tube-based formulations are relevant
here because prediction-error bounds can be incorporated through constraint
tightening while the online problem remains a quadratic program (QP). For
systems beyond the LTI class, comparable robust MPC results based on noisy
Hankel predictors are still largely unavailable. The LPV setting provides a
useful step in this direction, as it can describe nonlinear and time-varying
dynamics through scheduling-dependent linear models while retaining a
behavioral representation suitable for Hankel-based
prediction~\cite{Verhoek2026TAC_behavioral} and robust predictive control
analysis.

In this paper, we propose an LPV tube-based robust data-driven predictive
control (LPV-TRDDPC) scheme for unknown discrete-time LPV systems using offline
noisy input-scheduling-state multi-trajectory data. With noisy data,
unrestricted Hankel combinations generally do not admit a uniform bound on the
induced predictor mismatch, which is essential for tube-based tightening.
Imposing a scaled-simplex constraint yields such a bound, but may exclude
relevant constrained lifted trajectories from the corresponding Hankel
parametrization. This tension points to a more fundamental issue: for
constrained predictive control, it is sufficient to preserve the
finite-horizon trajectories relevant to the control task, rather than the
entire unconstrained behavior of the unknown LPV system.

The main contributions are as follows. First, we establish a multi-trajectory
simplified LPV fundamental lemma that combines several short
input-scheduling-state trajectories into a collective data-driven predictor,
avoiding the need for one long trajectory and increasing the flexibility of
data collection. Second, we introduce a constrained lifted behavior viewpoint
and combine a scaled-simplex Hankel parametrization with an offline convex-hull
coverage certificate for admissible one-step continuations of the optimal
control problem (OCP). This certificate preserves the representability
required by the receding-horizon argument while retaining a uniform noisy-data
predictor-mismatch bound. Third, we derive tube-tightening ingredients directly
from noisy data via LPV data-consistency sets and robust positively invariant
(RPI) sets. Finally, since the uncertainty and tightening ingredients are
computed offline, the receding-horizon controller is implemented online as a
QP without additional decision variables for uncertainty descriptions or tube
propagation. Under the stated scheduling-availability conditions and the
offline coverage certificate, recursive feasibility, closed-loop constraint
satisfaction, and practical input-to-state stability (ISS) are established.
To the best of our knowledge, this is the first tube-based direct data-driven
predictive control framework for LPV systems that establishes closed-loop
guarantees using noisy Hankel data. Since LTI systems are recovered as a
special case and LPV embeddings can capture certain classes of nonlinear
dynamics, the scope of the framework extends beyond the LTI setting. The
method is illustrated on an externally scheduled inverted oscillator.

The remainder of the paper is organized as follows. Section~\ref{sec:Preliminaries} introduces notation, formulates the problem, and develops the LPV data-driven trajectory representation together with the convex-hull coverage condition. Section~\ref{sec:Robust} presents the proposed LPV-TRDDPC scheme, derives the associated tightening construction, and establishes recursive feasibility, closed-loop constraint satisfaction, and practical ISS. Section~\ref{sec:examples} illustrates the proposed approach on an externally scheduled inverted oscillator. Section~\ref{sec:conclusions} concludes the paper.

\section{Preliminaries}\label{sec:Preliminaries}
This section collects notation and preliminaries. Section~\ref{subsec:ProblemSetup} states the problem of interest, and Section~\ref{sec:Data-driven} introduces the data-driven representation on which the later analysis and controller synthesis are built.
\subsection{Notation}\label{sec:notation}
Let $\mathbb{N}_0:=\{0,1,2,\ldots\}$ and, for integers $a\le b$, $\mathbb{N}_{a:b}:=\{a,a+1,\ldots,b\}$.
For a positive integer $n$, $I_n$ denotes the $n\times n$ identity matrix, and $0$ denotes a zero vector or matrix of compatible dimensions.
For a matrix $A$, $A^\top$ denotes its transpose, and
$\operatorname{vec}(A)$ denotes its column-wise vectorization.
For a vector $x$, $\|x\|=\|x\|_2$ denotes the Euclidean norm, and $\|x\|_\infty$ denotes the $\ell_\infty$ norm. For a matrix $A$, $\|A\|$ denotes the induced $2$-norm.
For $\mathcal A,\mathcal B\subset\mathbb{R}^n$, let $\mathcal A\oplus\mathcal B:=\{a+b\mid a\in\mathcal A,\ b\in\mathcal B\}$ and $\mathcal A\ominus\mathcal B:=\{x\in\mathbb{R}^n\mid \{x\}\oplus\mathcal B\subseteq\mathcal A\}$; for a matrix $M$, $M\mathcal A:=\{Ma\mid a\in\mathcal A\}$ denotes the linear image; for a matrix family $\mathcal M$, $\mathcal M\mathcal A:=\{Ma\mid M\in\mathcal M,\ a\in\mathcal A\}$; and $\operatorname{co}(\mathcal S)$, $\operatorname{cl}(\mathcal S)$, and $\otimes$ denote the convex hull, closure, and Kronecker product, respectively.
For vectors $a_0,\ldots,a_m$, let
$\operatorname{col}(a_0,\ldots,a_m):=[a_0^\top,\ldots,a_m^\top]^\top$.
For a matrix $M$, $\operatorname{cols}(M)$ denotes the set of its columns.
For a sequence $\{s_i\}_{i\in\mathbb Z}$ with
$s_i\in\mathbb R^{n_s}$, let
$s_{[i:j]}:=(s_i,\ldots,s_j)$ denote the finite segment from $i$ to $j$ and define
$s_{[i:j]}\odot I_n
:=\operatorname{blkdiag}(s_i\otimes I_n,\ldots,s_j\otimes I_n)$.
For absolute-time signals, let
$s_{[i,j]}:=\operatorname{col}(s_i,\ldots,s_j)$
denote the corresponding stacked vector. For a predicted signal $s$,
$s_{\ell|k}$ denotes its value predicted $\ell$ steps ahead at time $k$, and
$s_{[i|k,j|k]}:=\operatorname{col}(s_{i|k},\ldots,s_{j|k})$.

For a length-$T$ segment $s_{[0:T-1]}$, the depth-$L$ block Hankel matrix is
\[
H_L\!\big(s_{[0:T-1]}\big):=
\begin{bmatrix}
s_0 & s_1 & \cdots & s_{T-L}\\
s_1 & s_2 & \cdots & s_{T-L+1}\\
\vdots & \vdots & \ddots & \vdots\\
s_{L-1} & s_L & \cdots & s_{T-1}
\end{bmatrix}.
\]
For any Hankel matrix $H_s$ formed from an $n_s$-dimensional signal, its $\ell$th $n_s$-row block is denoted by $H_{s_{[\ell]}}$, and $H_{s_{[i:j]}}$ denotes the block-row submatrix formed by block rows $i,\ldots,j$.

\subsection{Problem Setup}\label{subsec:ProblemSetup}
We consider discrete-time LPV systems that can be represented by
state-space (SS) representations of the form:
\begin{subequations}\label{eq:lpv_ss}
\begin{align}
x_{k+1} &= A^\ast(p_k)x_k + B^\ast(p_k)u_k, \label{eq:lpv_ssa}\\
\hat{x}_k &= x_k + w_k,
\end{align}
\end{subequations}
where $x_k\in\mathbb{R}^{n_x}$, $u_k\in\mathbb{R}^{n_u}$, and
$p_k\in\mathbb{P}\subseteq\mathbb{R}^{n_p}$ are the state, input, and scheduling signals
at time instant $k\in\mathbb{Z}$, respectively. The true LPV system matrices depend affinely on the scheduling signal $p_k\in\mathbb P$ as
\begin{equation}\label{eq:affine_dep}
A^\ast(p_k)=A_0^\ast+\sum_{i=1}^{n_p} p_{k,i}A_i^\ast,\quad
B^\ast(p_k)=B_0^\ast+\sum_{i=1}^{n_p} p_{k,i}B_i^\ast,
\end{equation}
where $A_i^\ast\in\mathbb R^{n_x\times n_x}$ and $B_i^\ast\in\mathbb R^{n_x\times n_u}$ are unknown constant matrices. The affine parametrization in~\eqref{eq:affine_dep} is adopted for convenience and does not assign any special structural role to $A_0^\ast$ and $B_0^\ast$. The affine parametrization also includes polytopic LPV representations as a
special case, e.g., when \(A_0^\ast=B_0^\ast=0\) and \(p_k\) belongs to the unit simplex.
Throughout, \(\mathbb P\) is assumed to be compact and known. By reciprocal
rescaling of each scheduling coordinate and the corresponding matrices
\((A_i^\ast,B_i^\ast)\), known normalized bounds on \(\mathbb P\) can be imposed without
loss of generality. We assume that~\eqref{eq:lpv_ssa} is a nonredundant state
realization of the considered LPV behavior, in the sense that no state
coordinate can be removed without changing the corresponding
input-scheduling-state behavior. The data richness needed for the Hankel
representations is imposed separately through rank conditions.

The term $w_k\in\mathbb{R}^{n_x}$ denotes the measurement noise at time $k$ and satisfies $w_k\in\mathcal{W}$ for all $k$, where
\begin{equation}
\mathcal{W} := \{\, w \in \mathbb{R}^{n_x} \mid G_w w \le h_w \,\},
\label{eq:disturbance_polytope}
\end{equation}
is a known, compact, convex polytope satisfying either
$\mathcal W=\{0\}$ or $0\in\operatorname{int}\mathcal W$.

Our objective is to design a robust data-driven predictive control scheme for
the unknown LPV system~\eqref{eq:lpv_ss} using noisy data, while enforcing
closed-loop input and state constraints and stabilizing its equilibrium.

Both the input and the state are constrained to compact, convex polytopes
\begin{subequations}\label{eq:constraint_sets}
\begin{align}
\mathcal{U}
&=
\left\{\, u \in \mathbb{R}^{n_u} \;\middle|\; G_u u \leq h_u \,\right\},
\label{eq:constraint_sets_U}
\\
\mathcal{X}
&=
\left\{\, x \in \mathbb{R}^{n_x} \;\middle|\; G_x x \leq h_x \,\right\},
\label{eq:constraint_sets_X}
\end{align}
\end{subequations}
where $G_u,G_x$ and $h_u,h_x$ have compatible dimensions. We assume that $\mathcal U$ and $\mathcal X$ contain the
origin in their interiors.
The separated form is adopted to keep the tube-tightening notation transparent.
Joint polytopic state-input constraints in the combined variable $(x,u)$ can be
handled by tightening the corresponding inequalities.

\subsection{Data-driven Representation}\label{sec:Data-driven}
In analogy with the LTI case, data-driven controller synthesis requires a \emph{directly computable}
representation of the unknown dynamics from measured data. In the state-feedback setting considered here,
the LPV SS model~\eqref{eq:lpv_ssa} admits an equivalent shifted-affine LPV
input-output realization (with $y_k=x_k$). Hence, the simplified LPV Fundamental
Lemma~\cite{Verhoek2026TAC} can be applied to obtain an explicit Hankel matrix parametrization of LPV trajectories in terms of a coefficient
vector constructed from an input-scheduling-state dataset. In addition, to ensure the predictor is
well-defined over the receding horizon, we impose the standard scheduling availability assumption below.

\begin{assum}\label{as:scheduling_known}
At time step $k$, the scheduling trajectory $p_{[k,k+L-2]}$ is known up to time
step $k+L-2$.
\end{assum}
Accordingly, the preview used in the OCP is assumed to coincide with the
scheduling values subsequently realized by the plant. The present analysis
focuses on noisy offline data and online measurements. Extending it to scheduling prediction
errors, which enter through the lifted products and affect preview
compatibility, is a relevant direction for future work.

Next, we state and employ the simplified LPV Fundamental Lemma in input-state
form~\cite{Verhoek2026TAC}.
Superscript ${}^d$ denotes offline data, whereas superscript ${}^p$ denotes scheduling-lifted variables:
$u_k^p:=p_k\otimes u_k$, $x_k^p:=p_k\otimes x_k$, $\hat x_k^p:=p_k\otimes \hat x_k$, and $w_k^p:=p_k\otimes w_k$; the same convention is used for offline data, e.g., $u_k^{p,d}:=p_k^d\otimes u_k^d$ and $\hat x_k^{p,d}:=p_k^d\otimes \hat x_k^d$.

\begin{lem}\label{lem:FL_LPV_SS}
Consider the noise-free LPV system~\eqref{eq:lpv_ssa} and an input-scheduling-state sequence of state length $T$
$\bigl(u^d_{[0{:}T{-}2]},\allowbreak\,p^d_{[0{:}T{-}2]},\allowbreak\,x^d_{[0{:}T{-}1]}\bigr)$ from~\eqref{eq:lpv_ssa}.
Then, for any input-scheduling-state sequence
$\bigl(u_{[k,k+L-2]},\allowbreak\,p_{[k,k+L-2]},\allowbreak\,x_{[k,k+L-1]}\bigr)$ from~\eqref{eq:lpv_ssa}, there exists a vector
$g\in\mathbb R^{T-L+1}$ such that
\begin{equation}\label{eq:FL_LPV_SS}
\begin{bmatrix}
H_{L-1}\!\big(u^d_{[0{:}T{-}2]}\big)\\
H_{L}\!\big(x^d_{[0{:}T{-}1]}\big)\\
H_{L-1}\!\big(u^{p,d}_{[0{:}T{-}2]}\big)\!-\!\mathcal P_k^{n_u}\!H_{L-1}\!\big(u^d_{[0{:}T{-}2]}\big)\\
H_{L-1}\!\big(x^{p,d}_{[0{:}T{-}2]}\big)\!-\!\mathcal P_k^{n_x}\!H_{L-1}\!\big(x^d_{[0{:}T{-}2]}\big)
\end{bmatrix} \!g
=
\begin{bmatrix}
u_{[k,k+L-2]}\\
x_{[k,k+L-1]}\\
0\\
0
\end{bmatrix},
\end{equation}
where $\mathcal P^{\bullet}_k:=p_{[k,k+L-2]}\odot I_{\bullet}$, if and only if
\begin{equation}\label{eq:FL_LPV_SS_rank}
\operatorname{rank}\!\left(\!
\begin{bmatrix}
H_{L-1}\!\big(u^d_{[0{:}T{-}2]}\big)\\
H_{L}\!\big(x^d_{[0{:}T{-}1]}\big)\\
H_{L-1}\!\big(u^{p,d}_{[0{:}T{-}2]}\big)\\
H_{L-1}\!\big(x^{p,d}_{[0{:}T{-}2]}\big)
\end{bmatrix}\!
\right)
\!=\!
\bigl(n_p(n_x+n_u)+n_u\bigr)(L-1)+n_x.
\end{equation}
\end{lem}

Lemma~\ref{lem:FL_LPV_SS} is stated for a single offline input-scheduling-state trajectory. While this formulation is theoretically sufficient, it can be restrictive in practice, especially when additional design requirements must be imposed on the offline dataset. Using multiple measured trajectories offers substantially greater flexibility in data collection and selection. Motivated by the corresponding multiple-trajectory extension in the LTI case~\cite{vanWaarde2020CSL}, we therefore introduce a multi-trajectory simplified LPV Fundamental Lemma.

\begin{thm}
\label{thm:FL_LPV_SS_multi}
Consider the noise-free LPV system~\eqref{eq:lpv_ssa}. Let
\[
\bigl(u^{d,i}_{[0{:}T_i{-}2]},\allowbreak\,p^{d,i}_{[0{:}T_i{-}2]},\allowbreak\,x^{d,i}_{[0{:}T_i{-}1]}\bigr),
\qquad i\in\mathbb N_{1:q},
\]
be $q$ input-scheduling-state sequences of state lengths $T_i$ from~\eqref{eq:lpv_ssa}, where $T_i\ge L$ for all $i\in\mathbb N_{1:q}$. Define the collective Hankel matrices
\[
\begin{aligned}
H_u &:=
\big[\,H_{L-1}(u^{d,1}_{[0{:}T_1{-}2]})\ \cdots\ H_{L-1}(u^{d,q}_{[0{:}T_q{-}2]})\,\big],\\
H_x &:=
\big[\,H_{L}(x^{d,1}_{[0{:}T_1{-}1]})\ \cdots\ H_{L}(x^{d,q}_{[0{:}T_q{-}1]})\,\big],\\
H_{u^p} &:=
\big[\,H_{L-1}(u^{p,d,1}_{[0{:}T_1{-}2]})\ \cdots\ H_{L-1}(u^{p,d,q}_{[0{:}T_q{-}2]})\,\big],\\
H_{x^p} &:=
\big[\,H_{L-1}(x^{p,d,1}_{[0{:}T_1{-}2]})\ \cdots\ H_{L-1}(x^{p,d,q}_{[0{:}T_q{-}2]})\,\big].
\end{aligned}
\]
Then, for any state-horizon-$L$ input-scheduling-state sequence
$
\bigl(u_{[k,k+L-2]},\,p_{[k,k+L-2]},\,x_{[k,k+L-1]}\bigr)
$
from~\eqref{eq:lpv_ssa}, there exists a vector
$
g\in\mathbb R^{N_{\mathrm{col}}}$, where $
N_{\mathrm{col}}:=\sum_{i=1}^{q}(T_i-L+1)
$
such that
\begin{equation}
\label{eq:FL_LPV_SS_multi}
\begin{bmatrix}
H_u\\
H_x\\
H_{u^p}-\mathcal P_k^{n_u} H_u\\
H_{x^p}-\mathcal P_k^{n_x} H_{x_{[0:L-2]}}
\end{bmatrix} g
=
\begin{bmatrix}
u_{[k,k+L-2]}\\
x_{[k,k+L-1]}\\
0\\
0
\end{bmatrix},
\end{equation}
if and only if
\begin{equation}
\label{eq:FL_LPV_SS_multi_rank}
\operatorname{rank}\!\left(
\begin{bmatrix}
H_u\\
H_x\\
H_{u^p}\\
H_{x^p}
\end{bmatrix}
\right)
=
\bigl(n_p(n_x+n_u)+n_u\bigr)(L-1)+n_x.
\end{equation}
\end{thm}

\begin{pf}
The proof follows the same line as that of the single-trajectory case; see, e.g.,
Theorem~1 in~\cite{Verhoek2026TAC_behavioral}. The only modification is that the single Hankel matrix is
replaced by the collective Hankel matrix obtained by concatenating the corresponding
Hankel matrices of all measured trajectories. Hence, its column space is precisely
the span of all state-horizon-$L$ lifted subtrajectories extracted from the available data.
Moreover, the constraints
\[
\begin{bmatrix}
H_{u^p}-\mathcal P_k^{n_u} H_u\\[.2mm]
H_{x^p}-\mathcal P_k^{n_x} H_{x_{[0:L-2]}}
\end{bmatrix}g=0
\]
enforce consistency with the prescribed scheduling sequence
$p_{[k,k+L-2]}$. Therefore, the same argument as in the single-trajectory case
yields that every state-horizon-$L$ input-scheduling-state trajectory is
representable if and only if~\eqref{eq:FL_LPV_SS_multi_rank} holds.
\end{pf}

Theorem~\ref{thm:FL_LPV_SS_multi} shows that multiple short trajectories can be used in place of a single long one. In particular, each trajectory only needs to satisfy $T_i\ge L$ in order to contribute at least one valid state-horizon-$L$ column to the collective Hankel dictionary. When $T_i=L$, the $i$-th trajectory contributes exactly one column. For notational simplicity, we assume $T_i=L$ for all $i\in\mathbb N_{1:q}$ throughout the paper. Thus $N_{\mathrm{col}}=q$, and we write $g=\operatorname{col}(g_1,\ldots,g_q)$.

For controller synthesis, we assume that only $q$ offline measured input-scheduling-state trajectories
\[
\bigl(u^{d,i}_{[0{:}T_i{-}2]},\allowbreak\,p^{d,i}_{[0{:}T_i{-}2]},\allowbreak\,\hat x^{d,i}_{[0{:}T_i{-}1]}\bigr),
\qquad i\in\mathbb N_{1:q},
\]
from system~\eqref{eq:lpv_ss} are available. Fix a state depth $L$ and define
the \emph{collective} measured state and noise Hankel blocks
\begin{equation}
\label{eq:collective_Hankel_measured}
\begin{aligned}
H_{\hat x} &:=
\big[\,H_{L}(\hat x^{d,1}_{[0{:}T_1{-}1]})\ \cdots\ H_{L}(\hat x^{d,q}_{[0{:}T_q{-}1]})\,\big],\\
H_{\hat x^p} &:=
\big[\,H_{L-1}(\hat x^{p,d,1}_{[0{:}T_1{-}2]})\ \cdots\ H_{L-1}(\hat x^{p,d,q}_{[0{:}T_q{-}2]})\,\big],\\
H_w &:=
\big[\,H_{L}(w^{d,1}_{[0{:}T_1{-}1]})\ \cdots\ H_{L}(w^{d,q}_{[0{:}T_q{-}1]})\,\big],\\
H_{w^p} &:=
\big[\,H_{L-1}(w^{p,d,1}_{[0{:}T_1{-}2]})\ \cdots\ H_{L-1}(w^{p,d,q}_{[0{:}T_q{-}2]})\,\big],
\end{aligned}
\end{equation}
so that $H_{\hat x}=H_x+H_w$ and
$H_{\hat x^p}=H_{x^p}+H_{w^p}$.

In the noise-free setting, both Lemma~\ref{lem:FL_LPV_SS} and Theorem~\ref{thm:FL_LPV_SS_multi} provide an exact behavioral characterization of the unknown LPV dynamics through Hankel matrix equalities. In particular, the latter allows one to trade a single long trajectory for multiple short measured trajectories, provided that the collective richness condition~\eqref{eq:FL_LPV_SS_multi_rank} holds.

However, in the presence of measurement noise, the data-driven predictor inherits an effective multiplicative uncertainty, while the measurement-based initialization introduces an additional anchoring error. Under an unrestricted linear combination of Hankel columns, these mismatches may be arbitrarily amplified, so that no uniform offline prediction-error bound is available for ensuring closed-loop constraint satisfaction. Motivated by the tube-based robust data-driven MPC framework developed for the LTI case in~\cite{WangAngeli2026TRDDPC}, we therefore restrict the Hankel coefficient vector to the scaled simplex
\begin{equation}\label{eq:simplex_regularizer}
g\in\Delta^{N_{\mathrm{col}}},\qquad
\Delta^{N_{\mathrm{col}}}:=\{g\in\mathbb{R}^{N_{\mathrm{col}}}\mid g\ge 0,\ \mathbf{1}^\top g=\bar\alpha\},
\end{equation}
where $\mathbf{1}\in\mathbb{R}^{N_{\mathrm{col}}}$ denotes the all-ones vector and $\bar\alpha>0$ is a uniform prescribed scaling factor. Under~\eqref{eq:simplex_regularizer}, the mismatch accumulated by the data-driven predictor can be equivalently represented as a uniformly bounded blockwise additive residual acting on each predicted step, which in turn enables the polytopic tube tightening construction developed in Section~\ref{AA}.

The simplex restriction~\eqref{eq:simplex_regularizer}, however, introduces a new representability issue: although it regularizes the predictor and renders the induced uncertainty uniformly bounded, it may exclude valid constrained predicted trajectories that would otherwise be representable by an unrestricted coefficient vector. We first characterize scaled-convex-hull coverage of the noise-free constrained lifted behavior. This result will subsequently be specialized to the offline design target needed by the receding-horizon continuation. To this end, for the LPV system~\eqref{eq:lpv_ssa}, define the lifted variable $\zeta_k:=\operatorname{col}(u_k,x_k,u_k^p,x_k^p)$, where \(u_k^p:=p_k\otimes u_k\) and \(x_k^p:=p_k\otimes x_k\). Define also the block matrix
\begin{equation}
\label{eq:Theta_def}
\begin{aligned}
\Theta^\ast
&:=
\bigl[\,B_0^\ast\ \ A_0^\ast\ \ B_1^\ast\ \ \cdots\ \
B_{n_p}^\ast\ \ A_1^\ast\ \ \cdots\ \ A_{n_p}^\ast\,\bigr],
\\
\Theta^\ast&\in\mathbb R^{n_x\times n_\zeta},
\qquad
n_\zeta:=(n_p+1)(n_u+n_x).
\end{aligned}
\end{equation}
Then the LPV dynamics admit the lifted linear representation
\begin{equation}
\label{eq:lifted_linear_system_zeta}
x_{k+1}=\Theta^\ast\zeta_k.
\end{equation}
For a given horizon $L$, let
\[
\zeta_{[k,k+L-1]}
:=
\begin{bmatrix}
u_{[k,k+L-2]}\\
x_{[k,k+L-1]}\\
u^p_{[k,k+L-2]}\\
x^p_{[k,k+L-2]}
\end{bmatrix}
\]
denote the corresponding lifted state-horizon-$L$ trajectory.

Although the real LPV system~\eqref{eq:lpv_ssa} and the lifted
representation~\eqref{eq:lifted_linear_system_zeta} share the same state update
equation after lifting, their admissible lifted trajectories are different.
In the LPV system, the lifted signals are generated by a common scheduling
signal and hence satisfy the corresponding bilinear consistency relations. By
contrast, the lifted linear representation treats these signals as auxiliary
variables constrained only by the linear recursion. Hence, the lifted linear
behavior provides an outer embedding of the LPV behavior. We formalize this
distinction through the following finite-horizon constrained lifted sets.

Let
\[
\begin{aligned}
\mathcal U^p&:=\{p\otimes u\mid u\in\mathcal U,\ p\in\mathbb P\},\\
\mathcal X^p&:=\{p\otimes x\mid x\in\mathcal X,\ p\in\mathbb P\}.
\end{aligned}
\]
The constrained LPV lifted behavior is defined as
\begin{equation}
\label{eq:def_real_constrained_behavior_zeta}
\begin{aligned}
\mathfrak B_L^{C}
\!:=\!\bigl\{&\!
\operatorname{col}\bigl(
u_{[k,k+L-2]},x_{[k,k+L-1]},
u^p_{[k,k+L-2]},x^p_{[k,k+L-2]}
\bigr)\\
&\bigm|\,
\exists\,\{x_i\}_{i\in\mathbb N_{k:k+L-1}},\\
& \{u_i,p_i,u_i^p,x_i^p\}_{i\in\mathbb N_{k:k+L-2}}\ \text{s.t.}\\
& x_{i+1}=A^\ast(p_i)x_i+B^\ast(p_i)u_i,
\;i\in\mathbb N_{k:k+L-2},\\
& u_i\in\mathcal U,\quad i\in\mathbb N_{k:k+L-2},\\
& x_i\in\mathcal X,\quad i\in\mathbb N_{k:k+L-1},\\
& p_i\in\mathbb P,\quad u_i^p=p_i\otimes u_i,\\
& x_i^p=p_i\otimes x_i,\quad i\in\mathbb N_{k:k+L-2}
\bigr\}.
\end{aligned}
\end{equation}
Likewise, the constrained lifted linear outer behavior is defined as
\begin{equation}
\label{eq:def_lifted_constrained_behavior_zeta}
\begin{aligned}
\mathfrak B_L^{C'}
\!:=\!\bigl\{&\!
\operatorname{col}\bigl(
u_{[k,k+L-2]},x_{[k,k+L-1]},
v^u_{[k,k+L-2]},v^x_{[k,k+L-2]}
\bigr)\\
&\bigm|\,
\exists\,\{x_i\}_{i\in\mathbb N_{k:k+L-1}},\\
& \{u_i,v_i^u,v_i^x\}_{i\in\mathbb N_{k:k+L-2}}\ \text{s.t.}\\
& x_{i+1}=\Theta^\ast\!\operatorname{col}(u_i,x_i,v_i^u,v_i^x),
\;i\in\mathbb N_{k:k+L-2},\\
& u_i\in\mathcal U,\quad v_i^u\in\mathcal U^p,\\
& v_i^x\in\mathcal X^p,\quad i\in\mathbb N_{k:k+L-2},\\
& x_i\in\mathcal X,\quad i\in\mathbb N_{k:k+L-1}
\bigr\}.
\end{aligned}
\end{equation}
If \(\mathcal U,\mathcal X\), and \(\mathbb P\) are compact, then
\(\mathcal U^p\), \(\mathcal X^p\), \(\mathfrak B_L^C\), and
\(\mathfrak B_L^{C'}\) are compact. The set \(\mathfrak B_L^C\) enforces the
bilinear consistency
\[
(u_i^p,x_i^p)=(p_i\otimes u_i,p_i\otimes x_i),
\qquad i\in\mathbb N_{k:k+L-2},
\]
for a common scheduling value \(p_i\), whereas
\(\mathfrak B_L^{C'}\) treats \(v_i^u\) and \(v_i^x\) as auxiliary lifted variables
satisfying only the lifted variable bounds and the linear recursion. Hence,
\begin{equation}
\label{eq:BC_subset_BCp}
\mathfrak B_L^{C}\subseteq \mathfrak B_L^{C'}.
\end{equation}

Let
\[
H:=
\begin{bmatrix}
H_u\\
H_x\\
H_{u^p}\\
H_{x^p}
\end{bmatrix},
\]
where the collective Hankel matrices are formed from the corresponding
noise-free trajectories as in
Theorem~\ref{thm:FL_LPV_SS_multi}.

\begin{thm}
\label{thm:convex_hull_representation}
Suppose that
\[
\operatorname{rank}(H)
=
\bigl(n_p(n_x+n_u)+n_u\bigr)(L-1)+n_x.
\]
Moreover, assume that
\begin{equation}
\label{eq:origin_relative_interior_zeta}
0\in \operatorname{ri}_{\mathscr T_L}\!\bigl(\operatorname{co}(\operatorname{cols}(H))\bigr),
\qquad
\mathscr T_L:=\operatorname{span}(\operatorname{cols}(H)),
\end{equation}
where $\operatorname{ri}_{\mathscr T_L}(\cdot)$ denotes the relative interior in the
subspace $\mathscr T_L$. Then, there exists a finite scalar $\alpha>0$ such
that, for every $\zeta_{[k,k+L-1]}\in \mathfrak B_L^{C}$, there exists a vector
$g\in\mathbb R^{N_{\mathrm{col}}}$ such that
\begin{equation}
\label{eq:convex_hull_repr_reduced_zeta}
\begin{bmatrix}
H_u\\
H_x\\
H_{u^p}\!-\!\mathcal P_k^{n_u}\!H_u\\
H_{x^p}\!-\!\mathcal P_k^{n_x}\!H_{x_{[0:L-2]}}
\end{bmatrix}\!g
\!=\!
\begin{bmatrix}
u_{[k,k+L-2]}\\
x_{[k,k+L-1]}\\
0\\
0
\end{bmatrix},
g\ge 0,
\mathbf 1^\top g=\alpha.
\end{equation}
\end{thm}
\begin{pf}
By Theorem~\ref{thm:FL_LPV_SS_multi} and the rank condition, every length-$L$
trajectory of the lifted linear system~\eqref{eq:lifted_linear_system_zeta}
belongs to $\mathscr T_L=\operatorname{span}(\operatorname{cols}(H))$. Since
$\mathfrak B_L^{C}\subseteq \mathfrak B_L^{C'}$, any
$\zeta_{[k,k+L-1]}\in\mathfrak B_L^{C}$ therefore belongs to $\mathscr T_L$.
Fix $\zeta:=\zeta_{[k,k+L-1]}\in\mathfrak B_L^C$.

By~\eqref{eq:origin_relative_interior_zeta}, there exists $r>0$ such that
\[
\mathbb B_{\mathscr T_L}(0,r)
:=
\{\,\eta\in\mathscr T_L\mid \|\eta\|\le r\,\}
\subseteq
\operatorname{co}(\operatorname{cols}(H)).
\]
Since $0\in\operatorname{co}(\operatorname{cols}(H))$, there exists
$\lambda^0\in\mathbb R^{N_{\mathrm{col}}}$ such that
$H\lambda^0=0$, $\lambda^0\ge0$, and $\mathbf 1^\top\lambda^0=1$.
Moreover, since $\mathfrak B_L^C$ is compact, define
\[
M_C:=\max_{\xi\in\mathfrak B_L^C}\|\xi\|,
\qquad
\alpha:=\max\{1,M_C/r\}.
\]

If $\zeta\neq0$, set $\eta:=r\zeta/\|\zeta\|$.
Then $\eta\in\operatorname{co}(\operatorname{cols}(H))$, and hence there exists
$\lambda\in\mathbb R^{N_{\mathrm{col}}}$ satisfying
\[
H\lambda=\eta,\qquad \lambda\ge 0,\qquad \mathbf 1^\top\lambda=1.
\]
Set $\beta:=\|\zeta\|/r$ and
$g:=\beta\lambda+(\alpha-\beta)\lambda^0$.
Since $\beta\le M_C/r\le\alpha$, this vector is nonnegative. If
$\zeta=0$, set $g:=\alpha\lambda^0$. In both cases,
$g\ge0$, $\mathbf 1^\top g=\alpha$, and $Hg=\zeta$.

Finally, writing $Hg=\zeta$ blockwise and using the defining LPV
relations $u^p_{[k,k+L-2]}=\mathcal P_k^{n_u}u_{[k,k+L-2]}$ and
$x^p_{[k,k+L-2]}=\mathcal P_k^{n_x}x_{[k,k+L-2]}$ for
$\mathfrak B_L^{C}$ gives~\eqref{eq:convex_hull_repr_reduced_zeta}.
\end{pf}
The same relative-interior argument gives the following compact cover
consequence, used later for measured lifted prediction sets.

\begin{cor}
\label{cor:compact_cover_general}
Suppose that the relative interior
condition~\eqref{eq:origin_relative_interior_zeta} holds. Then, for any
nonempty compact set $\mathcal K\subseteq\mathscr T_L$, there exists a finite
scalar $\alpha_{\mathcal K}\ge 0$ such that
\begin{equation}
\label{eq:compact_cover_general}
\mathcal K
\subseteq
\alpha_{\mathcal K}
\operatorname{co}(\operatorname{cols}(H)).
\end{equation}
\end{cor}

\begin{pf}
Choose $r>0$ as in~\eqref{eq:origin_relative_interior_zeta}, so that
$\{\,\eta\in\mathscr T_L\mid \|\eta\|\le r\,\}
\subseteq\operatorname{co}(\operatorname{cols}(H))$.
Let $M_{\mathcal K}:=\max_{\zeta\in\mathcal K}\|\zeta\|$. If
$M_{\mathcal K}=0$, the claim holds with $\alpha_{\mathcal K}=0$. Otherwise,
for every $\zeta\in\mathcal K$, one has
$\frac{r}{M_{\mathcal K}}\zeta
\in\operatorname{co}(\operatorname{cols}(H))$ and hence
$\zeta\in\frac{M_{\mathcal K}}{r}
\operatorname{co}(\operatorname{cols}(H))$.
Thus~\eqref{eq:compact_cover_general} holds with
$\alpha_{\mathcal K}=M_{\mathcal K}/r$.
\end{pf}

\begin{rem}
\label{rem:alpha_outer_bound_zeta}
Theorem~\ref{thm:convex_hull_representation} concerns exact constrained LPV
trajectories in $\mathfrak B_L^C$ represented by the ideal Hankel dictionary
$H$. The corresponding geometric statement is given by
Corollary~\ref{cor:compact_cover_general}, which shows that the relative
interior condition guarantees a finite scaled-convex-hull cover for any
prescribed compact subset of $\mathscr T_L$.

The least scaling factor that would cover the ideal constrained behavior
depends on $\mathfrak B_L^C$, which is not explicitly available without the
true system. The scaling factor used in the controller is therefore selected as
a conservative offline design parameter.
\end{rem}

For later use in the robust OCP, we introduce the measured lifted Hankel
dictionary
\begin{equation}
\label{eq:measured_lifted_dictionary}
\hat H
:=
\begin{bmatrix}
H_u\\
H_{\hat x}\\
H_{u^p}\\
H_{\hat x^p}
\end{bmatrix},
\qquad
\hat{\mathscr T}_L
:=
\operatorname{span}\!\bigl(\operatorname{cols}(\hat H)\bigr).
\end{equation}
The corresponding measured-data relative interior condition is
\begin{equation}
\label{eq:origin_relative_interior_zhat}
0\in
\operatorname{ri}_{\hat{\mathscr T}_L}
\!\bigl(\operatorname{co}(\operatorname{cols}(\hat H))\bigr).
\end{equation}
If~\eqref{eq:origin_relative_interior_zhat} holds, then the same argument as in
Corollary~\ref{cor:compact_cover_general} implies that every compact set
$\hat{\mathcal K}\subseteq\hat{\mathscr T}_L$ admits a finite scaled-convex
cover; namely, there exists $\alpha_{\hat{\mathcal K}}\ge0$ such that
\begin{equation}
\label{eq:compact_cover_measured_dictionary}
\hat{\mathcal K}
\subseteq
\alpha_{\hat{\mathcal K}}
\operatorname{co}\!\bigl(\operatorname{cols}(\hat H)\bigr).
\end{equation}
This coverage property will be used below for admissible one-step
continuations of the robust OCP.

Since the system matrices are not available, we define equilibria directly in
terms of input-scheduling-state sequences.

\begin{defn}
\label{def:lpv_forced_equilibrium_ss}
A triple $(u_s,p_s,x_s)\in\mathbb R^{n_u}\times\mathbb P\times\mathbb R^{n_x}$ is a forced equilibrium of the
LPV system~\eqref{eq:lpv_ssa} if the constant sequence
$
(u_k,p_k,x_k)\equiv (u_s,p_s,x_s)
$
for all $k\in\mathbb N_0$ is a trajectory of~\eqref{eq:lpv_ssa}.
\end{defn}

In the remainder of the paper, we develop a robust data-driven predictive
control scheme built on Theorem~\ref{thm:FL_LPV_SS_multi},
Theorem~\ref{thm:convex_hull_representation}, and the
covering property in~\eqref{eq:compact_cover_measured_dictionary}. Throughout,
the ideal noise-free Hankel is used for the behavioral interpretation in
Theorem~\ref{thm:convex_hull_representation}, whereas the measured Hankel
$\hat H$ is used in the robust OCP and the continuation coverage condition.
The measured relative-interior and coverage conditions are offline,
a posteriori conditions on the collected data and can be certified by standard
convex-hull tests.
The equilibria are understood in the sense of
Definition~\ref{def:lpv_forced_equilibrium_ss}.

\section{LPV-TRDDPC}\label{sec:Robust}
In this section, we introduce the LPV-TRDDPC scheme under noisy data conditions. Section~\ref{Scheme} presents the LPV-TRDDPC scheme, while
Section~\ref{AA} details the tightening procedure.
Recursive feasibility, closed-loop constraint satisfaction, and practical ISS are
established in Section~\ref{sec:rec-feas}.

\subsection{Proposed Data-Driven Predictive Control Scheme}\label{Scheme}
Given the offline Hankel matrices $H_u$, $H_{\hat x}$, $H_{u^p}$, and $H_{\hat x^p}$ and the current measured
state $\hat x_k$, together with a scheduling preview
$\mathbf p_k:=p_{[0|k,L-2|k]}$, we define the LPV Hankel predictor
\begin{equation}\label{eq:lpv_hankel_predictor}
\begin{bmatrix}
u_{[0|k,L-2|k]}\\[.2em]
z_{[0|k,L-1|k]}\\[.2em]
0\\[.2em]
0
\end{bmatrix}
=
\begin{bmatrix}
H_u\\[.2em]
H_{\hat x}\\[.2em]
H_{u^p}-\mathcal P_k^{n_u}H_u\\[.2em]
H_{\hat x^p}-\mathcal P_k^{n_x}H_{\hat x_{[0:L-2]}}
\end{bmatrix} g.
\end{equation}
Here, $p_{0|k}=p_k$ and
$\mathcal P_k^\bullet:=p_{[0|k,L-2|k]}\odot I_\bullet$.

We then consider the following robust data-driven predictive OCP:
\begin{subequations}\label{eq:robust_deepc}
\begin{align}
\min_{\,z_{[0|k,L-1|k]},\,u_{[0|k,L-2|k]},\,g}\quad
& \sum_{\ell=0}^{L-2} J_{\mathrm s}\!\bigl(z_{\ell|k},u_{\ell|k}\bigr)+J_{\mathrm f}\!\bigl(z_{L-1|k}\bigr)
\label{eq:robust_deepc_obj}\\
\text{s.t.}\quad
& z_{0|k}-\hat x_k \in \mathcal E,
\label{eq:robust_deepc_init}\\
& \eqref{eq:simplex_regularizer}\ \text{holds},
\label{eq:robust_deepc_simplex}\\
& \eqref{eq:lpv_hankel_predictor}\ \text{holds},
\label{eq:robust_deepc_pred}\\
& z_{\ell|k}\in\mathcal Z,\qquad \forall\,\ell\in\mathbb N_{1:L-1},
\label{eq:robust_deepc_state}\\
& u_{\ell|k}\in\hat{\mathcal U},\qquad \forall\,\ell\in\mathbb N_{0:L-2},
\label{eq:robust_deepc_input}\\
& z_{L-1|k}\in \mathcal Z_{\mathrm f}.
\label{eq:robust_deepc_terminal}
\end{align}
\end{subequations}
Here, $\mathcal Z\subset\mathbb R^{n_x}$ and $\hat{\mathcal U}\subset\mathbb R^{n_u}$ denote the tightened state and input polytopes, respectively. In Section~\ref{sec:rec-feas}, $\mathcal E$ will be synthesized as an RPI polytope for the anchor-error recursion, and a terminal set $\mathcal Z_{\mathrm f}\subseteq\mathcal Z$ will be selected as an RPI polytope for the terminal feedback law. Since OCP~\eqref{eq:robust_deepc} is parameterized by the scheduling preview, we denote the corresponding measured- and true-state feasible sets by
\begin{equation}
\label{eq:feasible_sets_def}
\begin{aligned}
\hat{\mathcal X}^{\mathrm{feas}}_{L}(\mathbf p)
&:=
\bigl\{\hat x\in\mathbb R^{n_x}\bigm|
\text{OCP~\eqref{eq:robust_deepc} is feasible at }(\hat x,\mathbf p)\bigr\},
\\
\mathcal X^{\mathrm{feas}}_{L}(\mathbf p)
&:=
\hat{\mathcal X}^{\mathrm{feas}}_{L}(\mathbf p)\ominus\mathcal W.
\end{aligned}
\end{equation}
Denote an optimizer of~\eqref{eq:robust_deepc} by
$\bigl(z^\star_{[0|k,L-1|k]},\allowbreak\,u^\star_{[0|k,L-2|k]},\allowbreak\,g^\star\bigr)$.
The stage and terminal costs are quadratic with respect to the prescribed
equilibrium pair $(x_s,u_s)$ introduced in
Definition~\ref{def:lpv_forced_equilibrium_ss}:
\begin{equation}\label{eq:QR_costs}
\begin{aligned}
J_{\mathrm{s}}\!\bigl(z_{\ell|k}, u_{\ell|k}\bigr)
&=(z_{\ell|k}-x_s)^\top Q(z_{\ell|k}-x_s)\\
&\quad +(u_{\ell|k}-u_s)^\top R(u_{\ell|k}-u_s),\\
J_{\mathrm f}(z_{L-1|k})
&=(z_{L-1|k}-x_s)^\top P_L(z_{L-1|k}-x_s),
\end{aligned}
\end{equation}
with $Q\succeq \underline{q}I \succ 0$, $R\succeq \underline{r}I \succ 0$, and
$P_{L}\succ0$; hence, the OCP~\eqref{eq:robust_deepc} is a QP.
At each time $k$, after solving OCP~\eqref{eq:robust_deepc}, the first optimized
nominal input is applied with the feedback correction
\begin{equation}
\label{eq:kappa_deepc_clean}
u_k
=\kappa_L^\star\bigl(\hat x_k,\mathbf p_k\bigr):=u_{0|k}^\star+K\bigl(\hat x_k-z_{0|k}^\star\bigr),
\end{equation}
where $K$ is a scheduling-independent common stabilizing gain computed offline
from noisy data through Proposition~\ref{prop:lossless_grid_vertices}. It does
not enter the Hankel predictor~\eqref{eq:lpv_hankel_predictor}; hence,
$z_{[0|k,L-1|k]}$ is an open-loop nominal prediction generated by the optimized
input sequence. Instead, $K$ serves as an ancillary tube and terminal gain: it
corrects the first implemented input through~\eqref{eq:kappa_deepc_clean} and
induces the closed-loop family used to propagate the anchoring error and to
construct the RPI, tightening, and terminal
ingredients. The resulting robust data-driven predictive control scheme is
summarized in Algorithm~\ref{alg:TRDDPC}.
\begin{algorithm}[tbp]
  \caption{LPV-TRDDPC}
  \label{alg:TRDDPC}
\begin{algorithmic}[1]
    \STATE Measure the current noisy state $\hat x_k$ at time step $k$.
    \STATE Obtain the scheduling preview $\mathbf p_k=p_{[0|k,L-2|k]}$.
    \STATE Solve the OCP~\eqref{eq:robust_deepc}.
    \STATE Apply $u_k = \kappa_L^\star(\hat x_k,\mathbf p_k)$.
    \STATE Set $k \gets k{+}1$ and return to Step~1.
  \end{algorithmic}
\end{algorithm}

\subsection{Constraint Tightening}\label{AA}
To ensure that the \emph{actual} closed-loop state and input satisfy the original constraints, we tighten the
constraints by explicitly bounding the prediction error. We next show how the tightened sets $\mathcal Z$ and
$\hat{\mathcal U}$ can be constructed \emph{directly from noisy data}. Consider any prediction
$\bigl(z_{[0|k,L-1|k]},u_{[0|k,L-2|k]},g\bigr)$
satisfying the OCP constraints in~\eqref{eq:robust_deepc}. The following
auxiliary closed-loop sequences are introduced only for the error analysis. For an admissible
measurement noise sequence $w_{k+\ell}\in\mathcal W$, let
$\hat x_{\ell|k}:=x_{\ell|k}+w_{k+\ell}$ and define the auxiliary sequences
\begin{subequations}\label{eq:aux_sequences_tube}
\begin{align}
\bar x_{0|k}&:=z_{0|k},\label{eq:barx_dynamics_main}\\
\bar x_{\ell+1|k}
&:=A^\ast(p_{\ell|k})\,z_{\ell|k}
+B^\ast(p_{\ell|k})\,u_{\ell|k},
\nonumber\\
\tilde x_{0|k}&:=\hat x_k,\label{eq:tildex_dynamics_main}\\
\tilde x_{\ell+1|k}
&:=A^\ast\!(p_{\ell|k})\hat x_{\ell|k}
+B^\ast\!(p_{\ell|k})
\bigl(u_{\ell|k}+K\!(\hat x_{\ell|k}\!-z_{\ell|k})\bigr),
\nonumber\\
x_{0|k}&:=x_k,\label{eq:x_dynamics_main}\\
x_{\ell+1|k}
&:=A^\ast\!(p_{\ell|k})x_{\ell|k}
+B^\ast\!(p_{\ell|k})
\bigl(u_{\ell|k}+K\!(\hat x_{\ell|k}\!-z_{\ell|k})\bigr).
\nonumber
\end{align}
\end{subequations}
For these sequences, the prediction error admits the decomposition
\begin{equation}\label{eq:tube_chain_intro}
z_{\ell|k}-x_{\ell|k}
=
(z_{\ell|k}-\bar x_{\ell|k})
+(\bar x_{\ell|k}-\tilde x_{\ell|k})
+(\tilde x_{\ell|k}-x_{\ell|k}).
\end{equation}
The three terms in~\eqref{eq:tube_chain_intro} isolate, respectively, the predictor (offline) mismatch, the
propagation of the anchoring error $z_{0|k}-\hat x_k\in\mathcal E$, and the contribution of the measurement noise
$w_{k+\ell}\in\mathcal W$. Each term admits a compact polytopic bound; consequently, Minkowski summation yields an
explicit polytopic bound on $z_{\ell|k}-x_{\ell|k}$ (and the induced input deviation), which leads to the
tightened constraints in~\eqref{eq:robust_deepc_state}--\eqref{eq:robust_deepc_input}.

To obtain computable bounds from noisy data, we introduce LPV one-step
consistency sets for candidate affine models. Let
$n_\theta:=n_x n_\zeta$. For any $\theta\in\mathbb R^{n_\theta}$, let
$\Theta_\theta:=\operatorname{mat}_{n_x,n_\zeta}(\theta)$, where
$\operatorname{mat}_{n_x,n_\zeta}$ is the inverse of $\operatorname{vec}$, and
partition
\begin{equation}\label{eq:theta_matrix_partition}
\begin{aligned}
\Theta_\theta
&:=
\bigl[\,B_{\theta,0}\ \ A_{\theta,0}\ \ B_{\theta,1}
\  \cdots\ \
B_{\theta,n_p}\ \ A_{\theta,1}\  \cdots\ \ A_{\theta,n_p}\,\bigr],
\end{aligned}
\end{equation}
where $A_{\theta,j}\in\mathbb R^{n_x\times n_x}$ and
$B_{\theta,j}\in\mathbb R^{n_x\times n_u}$ for all
$j\in\mathbb N_{0:n_p}$. The LPV matrices induced by $\theta$ are
\begin{equation}\label{eq:A_theta_B_theta_def}
\begin{aligned}
A_\theta(p)&:=A_{\theta,0}+\sum_{j=1}^{n_p}p_jA_{\theta,j},
\\
B_\theta(p)&:=B_{\theta,0}+\sum_{j=1}^{n_p}p_jB_{\theta,j},
\end{aligned}
\end{equation}
so that $A_\theta(p)\in\mathbb R^{n_x\times n_x}$ and
$B_\theta(p)\in\mathbb R^{n_x\times n_u}$ for all $p\in\mathbb P$. The true
system corresponds to $\theta^\ast:=\operatorname{vec}(\Theta^\ast)$.

For a compact convex set $\mathcal W$ containing the origin and a matrix $A$,
we use the $\mathcal W$-set gain
$\|A\|_{\mathcal W}:=\inf\{\rho\ge0\mid
A\mathcal W\subseteq\rho\mathcal W\}$.
This quantity is used only as a set-inclusion gain; it need not be a norm when
$\mathcal W$ is not centrally symmetric.

For the offline measured multi-trajectory dataset
$\bigl(u^{d,i}_{[0{:}T_i{-}2]},\allowbreak\,p^{d,i}_{[0{:}T_i{-}2]},\allowbreak\,
\hat x^{d,i}_{[0{:}T_i{-}1]}\bigr)$, let
$\mathscr J:=\mathbb N_{1:q}\times\mathbb N_{0:L-2}$. For
$(i,\ell)\in\mathscr J$ and $\gamma_{i,\ell}\ge0$, define the inflated
disturbance-difference set
$\mathcal D(\gamma_{i,\ell}):=\mathcal W\oplus
\bigl(-\gamma_{i,\ell}\mathcal W\bigr)$.
The set \(\mathcal D(\gamma_{i,\ell})\) specifies the admissible one-step
fitting residual used in the consistency test. In particular, for the true
parameter \(\theta^\ast\), the disturbance-difference term
$
w^{d,i}_{\ell+1}
-
A^\ast(p^{d,i}_{\ell})w^{d,i}_{\ell}
$
belongs to \(\mathcal D(\gamma_{i,\ell})\) whenever
$
A^\ast(p^{d,i}_{\ell})\mathcal W
\subseteq
\gamma_{i,\ell}\mathcal W$, or equivalently
\(\|A^\ast(p^{d,i}_{\ell})\|_{\mathcal W}\le\gamma_{i,\ell}\) in the set-gain sense. We then define
the one-step LPV consistency set
\begin{equation}\label{eq:Ci_theta_gammai}
\mathcal C_{i,\ell}^{\theta}(\gamma_{i,\ell})
:=
\left\{\theta\,\middle|\,
\begin{aligned}
\exists\,d_{i,\ell} &\in \mathcal D(\gamma_{i,\ell}):\\
\hat x_{\ell+1}^{d,i}
&= A_\theta(p_{\ell}^{d,i})\hat x_{\ell}^{d,i}\\
 &\quad+ B_\theta(p_{\ell}^{d,i})u_{\ell}^{d,i}
 + d_{i,\ell}
\end{aligned}
\right\}.
\end{equation}
The corresponding data-consistency set in $\theta$-space is
\begin{equation}\label{eq:I_theta_vector_gamma}
\mathcal I^\theta(\boldsymbol\gamma)
:=
\bigcap_{(i,\ell)\in\mathscr J}
\mathcal C_{i,\ell}^{\theta}(\gamma_{i,\ell}),
\qquad
\boldsymbol\gamma:=
(\gamma_{i,\ell})_{(i,\ell)\in\mathscr J}.
\end{equation}
For each $\ell\in\mathbb N_{0:L-2}$, we further define the window data-consistency set
\begin{equation}\label{eq:I_set_theta_window}
\mathcal I^\theta_\ell(\boldsymbol\gamma_\ell)
:=
\bigcap_{i=1}^{q}\mathcal C_{i,\ell}^{\theta}(\gamma_{i,\ell}),
\qquad
\boldsymbol\gamma_\ell:=(\gamma_{1,\ell},\ldots,\gamma_{q,\ell}).
\end{equation}
For polytopic $\mathcal W$, the sets
$\mathcal C_{i,\ell}^{\theta}(\gamma_{i,\ell})$ and their finite
intersections are polyhedra.

We next show that, with respect to the above multi-trajectory dataset, every feasible Hankel-predicted trajectory admits a one-step realization under some stepwise data-consistent LPV coefficient $\theta_\ell\in\mathcal I^\theta_\ell(\boldsymbol\gamma_\ell)$, up to an additive residual induced by the one-step consistency slacks.
\begin{prop}\label{prop:window_model_realization_TAC_poly_LPV}
Fix $\ell\in\mathbb N_{0:L-2}$ and let $\theta_\ell\in\mathcal I^\theta_\ell(\boldsymbol\gamma_\ell)$.
Let $g\in\mathbb R^{N_{\mathrm{col}}}$ satisfy~\eqref{eq:robust_deepc_simplex}--\eqref{eq:robust_deepc_pred}.
Define
\[
\mathcal R_\ell(\boldsymbol\gamma_\ell)
:=
\bar\alpha\,
\operatorname{co}\Bigl(
\bigcup_{i=1}^{q}\mathcal D(\gamma_{i,\ell})
\Bigr).
\]
Then each Hankel-predicted step admits the one-step realization
\begin{equation}\label{eq:z_dyn_window_TAC_poly_LPV}
z_{\ell+1|k}
=
A_{\theta_\ell}(p_{\ell|k})\,z_{\ell|k}
+
B_{\theta_\ell}(p_{\ell|k})\,u_{\ell|k}
+
\Delta_\ell(g),
\end{equation}
where $\Delta_\ell(g)\in\mathcal R_\ell(\boldsymbol\gamma_\ell)$.
\end{prop}

\begin{pf}
Fix $\ell$ and $\theta_\ell\in\mathcal I^\theta_\ell(\boldsymbol\gamma_\ell)$.
Since $\theta_\ell\in\mathcal C_{i,\ell}^{\theta}(\gamma_{i,\ell})$ for all $i\in\mathbb N_{1:q}$, for each such $i$
there exists $d_{i,\ell}\in\mathcal D(\gamma_{i,\ell})$ such that
\begin{equation}\label{eq:one_step_TAC_poly_proof_LPV_short}
\hat x^{d,i}_{\ell+1}
=
A_{\theta_\ell}(p^{d,i}_{\ell})\hat x^{d,i}_{\ell}
+
B_{\theta_\ell}(p^{d,i}_{\ell})u^{d,i}_{\ell}
+
d_{i,\ell}.
\end{equation}
Multiplying by $g_i$ and summing over $i\in\mathbb N_{1:q}$ yields
\begin{equation}\label{eq:window_stack_short}
H_{\hat x_{[\ell+1]}}g
=
\sum_{i=1}^{q}g_i
\Bigl(
A_{\theta_\ell}(p^{d,i}_{\ell})\hat x^{d,i}_{\ell}
+
B_{\theta_\ell}(p^{d,i}_{\ell})u^{d,i}_{\ell}
+d_{i,\ell}
\Bigr).
\end{equation}
Using~\eqref{eq:robust_deepc_pred}, $z_{\ell|k}=H_{\hat x_{[\ell]}}g$, $z_{\ell+1|k}=H_{\hat x_{[\ell+1]}}g$, and
$u_{\ell|k}=H_{u_{[\ell]}}g$. Moreover, the LPV predictor constraints imply
\[
H_{u^p_{[\ell]}}g=(p_{\ell|k}\otimes I_{n_u})H_{u_{[\ell]}}g,\qquad
H_{\hat x^p_{[\ell]}}g=(p_{\ell|k}\otimes I_{n_x})H_{\hat x_{[\ell]}}g.
\]
Together with the affine dependence of $A_{\theta_\ell}(\cdot)$ and $B_{\theta_\ell}(\cdot)$, this yields
\[\begin{aligned}
\sum_{i=1}^{q}g_i A_{\theta_\ell}(p^{d,i}_{\ell})\hat x^{d,i}_{\ell}
=
A_{\theta_\ell}(p_{\ell|k})z_{\ell|k},\\
\sum_{i=1}^{q}g_i B_{\theta_\ell}(p^{d,i}_{\ell})u^{d,i}_{\ell}
=
B_{\theta_\ell}(p_{\ell|k})u_{\ell|k}.
\end{aligned}
\]
Substituting into~\eqref{eq:window_stack_short}
gives~\eqref{eq:z_dyn_window_TAC_poly_LPV} with
$\Delta_\ell(g):=\sum_{i=1}^{q}g_i d_{i,\ell}$.
Since $g\ge0$, $\mathbf 1^\top g=\bar\alpha$, and
$d_{i,\ell}\in\mathcal D(\gamma_{i,\ell})$, it follows that
$\Delta_\ell(g)\in\mathcal R_\ell(\boldsymbol\gamma_\ell)$.
\end{pf}
By Proposition~\ref{prop:window_model_realization_TAC_poly_LPV}, each
predicted step admits a data-consistent one-step realization with a bounded
residual. We next construct a compact initial parameter enclosure directly from
the noisy data. Define
$\hat\zeta_{\ell}^{d,i}:=\operatorname{col}(u^{d,i}_{\ell},
\allowbreak\,\hat x^{d,i}_{\ell},\allowbreak\,u^{p,d,i}_{\ell},
\allowbreak\,\hat x^{p,d,i}_{\ell})$.
Using a fixed ordering of $\mathscr J$, set
\begin{equation}\label{eq:stacked_lifted_regression_data}
\begin{aligned}
\hat Z_\zeta&:=
\bigl[\,\hat\zeta_{\ell}^{d,i}\,\bigr]_{(i,\ell)\in\mathscr J},\\
\hat X_+&:=
\bigl[\,\hat x_{\ell+1}^{d,i}\,\bigr]_{(i,\ell)\in\mathscr J}.
\end{aligned}
\end{equation}
For an admissible offline noise sequence
$\mathbf w^d\in\mathcal W^{qL}$, define
$e_{\ell}^{d,i}(\mathbf w^d):=\operatorname{col}(0,\allowbreak\,
w_{\ell}^{d,i},\allowbreak\,0,\allowbreak\,
p_{\ell}^{d,i}\otimes w_{\ell}^{d,i})$
and
\begin{equation}\label{eq:stacked_lifted_noise_data}
\begin{aligned}
E_\zeta(\mathbf w^d)
&:=\bigl[\,e_{\ell}^{d,i}(\mathbf w^d)\,\bigr]_{(i,\ell)\in\mathscr J},\\
W_+(\mathbf w^d)
&:=\bigl[\,w_{\ell+1}^{d,i}\,\bigr]_{(i,\ell)\in\mathscr J}.
\end{aligned}
\end{equation}
Choose finite bounds $\delta_\zeta,\delta_+\ge0$ such that
\begin{equation}\label{eq:stacked_noise_bounds}
\|E_\zeta(\mathbf w^d)\|_2\le\delta_\zeta,
\qquad
\|W_+(\mathbf w^d)\|_2\le\delta_+,
\end{equation}
for every $\mathbf w^d\in\mathcal W^{qL}$.
For example, if
$\epsilon:=\max_{w\in\mathcal W}\|w\|_2$, then
the choices
\begin{equation}\label{eq:explicit_stacked_noise_bounds}
\begin{aligned}
\delta_\zeta
&=\epsilon
\biggl(\sum_{(i,\ell)\in\mathscr J}
\bigl(1+\|p_\ell^{d,i}\|_2^2\bigr)\biggr)^{1/2},\\
\delta_+
&=\epsilon\sqrt{q(L-1)}
\end{aligned}
\end{equation}
satisfy~\eqref{eq:stacked_noise_bounds}.

\begin{assum}\label{as:robust_lifted_informativity}
The number of one-step samples satisfies $q(L-1)\ge n_\zeta$, and the measured
lifted regressor satisfies
\begin{equation}\label{eq:robust_lifted_rank}
\mu_\zeta
:=
\sigma_{\min}(\hat Z_\zeta)-\delta_\zeta
>0.
\end{equation}
\end{assum}
Condition~\eqref{eq:robust_lifted_rank} is a quantitative counterpart of a
full-row-rank test: every noise-free lifted regressor compatible with
$\mathcal W$ has full row rank. Minimum singular value conditions of this type
are used to quantify data richness in the robust Fundamental
Lemma~\cite{Coulson2023CSL}.

\begin{prop}\label{prop:initial_model_enclosure}
Suppose Assumption~\ref{as:robust_lifted_informativity} holds and define
\begin{equation}\label{eq:initial_model_enclosure_constants}
\begin{aligned}
\hat\Theta&:=\hat X_+\hat Z_\zeta^\dagger,
&\hat\theta&:=\operatorname{vec}(\hat\Theta),\\
\bar M_\Theta
&:=\frac{\|\hat X_+\|_2+\delta_+}{\mu_\zeta},
&r_\Theta
&:=\frac{\delta_++\bar M_\Theta\delta_\zeta}
{\sigma_{\min}(\hat Z_\zeta)}.
\end{aligned}
\end{equation}
Then the compact polytope
\begin{equation}\label{eq:initial_model_enclosure}
\mathcal I_0^\theta
:=
\bigl\{\theta\in\mathbb R^{n_\theta}
\bigm|\|\theta-\hat\theta\|_\infty\le r_\Theta\bigr\}
\end{equation}
contains the true parameter $\theta^\ast$.
\end{prop}

\begin{pf}
Let $E_\zeta^\ast$ and $W_+^\ast$ be the matrices
in~\eqref{eq:stacked_lifted_noise_data} associated with the realized offline
noise, and set
$Z_\zeta^\ast:=\hat Z_\zeta-E_\zeta^\ast$ and
$X_+^\ast:=\hat X_+-W_+^\ast$. By~\eqref{eq:robust_lifted_rank} and the
singular-value perturbation bound,
$\sigma_{\min}(Z_\zeta^\ast)\ge\mu_\zeta>0$. Since
$X_+^\ast=\Theta^\ast Z_\zeta^\ast$, it follows that
$\|\Theta^\ast\|_2\le\bar M_\Theta$. Moreover,
$\hat X_+=\Theta^\ast\hat Z_\zeta+W_+^\ast-\Theta^\ast E_\zeta^\ast$.
As $\hat Z_\zeta$ has full row rank,
$\hat\Theta-\Theta^\ast
=(W_+^\ast-\Theta^\ast E_\zeta^\ast)\hat Z_\zeta^\dagger$,
and hence $\|\hat\Theta-\Theta^\ast\|_2\le r_\Theta$. The claim follows
from $\|\operatorname{vec}(M)\|_\infty\le\|M\|_2$.
\end{pf}

Starting from $\mathcal I_0^\theta$, the finite nested refinement in
Appendix~\ref{app:gamma-tight-LPV} defines an inflation vector
$\boldsymbol\gamma^\star$ and a compact polytope
$\mathcal I_\star^\theta$ satisfying
\begin{equation}\label{eq:refined_model_set_properties}
\theta^\ast\in\mathcal I_\star^\theta
\subseteq\mathcal I^\theta(\boldsymbol\gamma^\star).
\end{equation}
The refinement tightens the initial data-consistent enclosure and can thereby
reduce conservatism in the model family and mismatch bounds used below.
We then choose a compact convex, preferably polytopic, set $\mathcal W_A$
satisfying
\begin{equation}\label{eq:WA_star_uniform_LPV}
A_\theta(p)\mathcal W
\subseteq
\mathcal W_A,
\qquad
\forall\,(\theta,p)\in\mathcal I_\star^\theta\times\mathbb P.
\end{equation}
One admissible scalar choice is $\mathcal W_A=\gamma^\star\mathcal W$, where
\[
\gamma^\star
\ \ge\
\sup_{p\in\mathbb P}\ \sup_{\theta\in\mathcal I_\star^\theta}
\|A_\theta(p)\|_{\mathcal W}.
\]
For a centrally symmetric disturbance polytope, this scalar choice gives
$\mathcal W\oplus(-\mathcal W_A)=(1+\gamma^\star)\mathcal W$. In general, however, any set satisfying~\eqref{eq:WA_star_uniform_LPV} may be used; this permits componentwise or direction-dependent bounds when these provide a tighter outer approximation. We denote
\begin{equation}\label{eq:DA_star_def_LPV}
\mathcal D_A
:=
\mathcal W\oplus(-\mathcal W_A).
\end{equation}
With $\theta_\ell=\theta^\ast$, the residual in
Proposition~\ref{prop:window_model_realization_TAC_poly_LPV} is generated by
$d_{i,\ell}^\ast:=w_{\ell+1}^{d,i}-A^\ast(p_\ell^{d,i})w_\ell^{d,i}$.
Since $w_{\ell+1}^{d,i}\in\mathcal W$ and
$A^\ast(p_\ell^{d,i})w_\ell^{d,i}\in\mathcal W_A$ by~\eqref{eq:WA_star_uniform_LPV}, one has
$d_{i,\ell}^\ast\in\mathcal D_A$. Hence, for any
$g\in\Delta^{N_{\mathrm{col}}}$,
$\Delta_\ell^\ast(g):=\sum_{i=1}^{q}g_i d_{i,\ell}^\ast
\in\bar\alpha\,\mathcal D_A$.
Accordingly, define the associated set of frozen data-consistent LPV matrix
pairs
\begin{equation}\label{eq:M_def_LPV}
\mathcal M
:=
\bigl\{(A_\theta(p),B_\theta(p))
\ \big|\
\theta\in\mathcal I_\star^\theta,\ p\in\mathbb P
\bigr\}.
\end{equation}

The corresponding closed-loop matrix sets induced by the common stabilizing
gain $K$ are
\begin{equation}\label{eq:Acl_set_def}
\begin{aligned}
\mathcal A_{\mathrm{cl}}(p)
&:=
\bigl\{A_\theta(p)+B_\theta(p)K
\ \big|\ \theta\in\mathcal I_\star^\theta\bigr\},\\
\mathcal A_{\mathrm{cl}}
&:=
\bigcup_{p\in\mathbb P}\mathcal A_{\mathrm{cl}}(p).
\end{aligned}
\end{equation}
When $\mathcal M$ is available through a polytopic outer approximation with vertices $
\operatorname{vert}(\mathcal M)=\{(A^{(j)},B^{(j)})\}_{j=1}^{N_v},
$
the induced closed-loop set admits the outer approximation
\begin{equation}\label{eq:Acl_poly_outer}
\mathcal A_{\mathrm{cl}}
\ \subseteq\
\operatorname{co}\bigl\{A_{\mathrm{cl}}^{(j)}\bigr\}_{j=1}^{N_v},
\qquad
A_{\mathrm{cl}}^{(j)}:=A^{(j)}+B^{(j)}K.
\end{equation}
Hence, in the sequel we work with the vertex set
$
\operatorname{vert}(\mathcal A_{\mathrm{cl}})=\{A_{\mathrm{cl}}^{(j)}\}_{j=1}^{N_v},
$
understood as a polytopic outer approximation whenever~\eqref{eq:Acl_poly_outer} is strict.
By~\eqref{eq:refined_model_set_properties},
$\bigl(A^\ast(p),B^\ast(p)\bigr)\in\mathcal M$ for all $p\in\mathbb P$.
We now define the error bounds
\begin{equation}\label{eq:error_sets_def_LPV}
\mathcal E_w:=\mathcal W_A,
\mathcal E_{\mathrm{on}}:=\operatorname{co}\bigl(\mathcal A_{\mathrm{cl}}\mathcal E\bigr),
\mathcal E_{\mathrm{off}}:=\bar\alpha\,\mathcal D_A.
\end{equation}
With these definitions, the three components in~\eqref{eq:tube_chain_intro} satisfy
\begin{subequations}\label{eq:three_step_bounds_LPV}
\begin{align}
\tilde x_{\ell|k}-x_{\ell|k}
&\in
\mathcal E_w,
&& \forall\,\ell\in\mathbb N_{1:L-1},
\label{eq:three_step_bounds_LPV_w}
\\
\bar x_{\ell|k}-\tilde x_{\ell|k}
&\in
\mathcal E_{\mathrm{on}},
&& \forall\,\ell\in\mathbb N_{1:L-1},
\label{eq:three_step_bounds_LPV_on}
\\
z_{\ell|k}-\bar x_{\ell|k}
&\in
\mathcal E_{\mathrm{off}},
&& \forall\,\ell\in\mathbb N_{1:L-1}.
\label{eq:three_step_bounds_LPV_off}
\end{align}
\end{subequations}
Here,~\eqref{eq:three_step_bounds_LPV_w} follows from the measurement noise propagated through the true LPV dynamics and the image bound~\eqref{eq:WA_star_uniform_LPV};~\eqref{eq:three_step_bounds_LPV_on} follows from the invariance of $\mathcal E$ under the robust closed-loop family $\mathcal A_{\mathrm{cl}}$; and~\eqref{eq:three_step_bounds_LPV_off} follows from Proposition~\ref{prop:window_model_realization_TAC_poly_LPV} specialized to $\boldsymbol\gamma^\star$, which yields the uniform offline mismatch bound $\bar\alpha\,\mathcal D_A$ compatible with the measured dataset. By construction,
\[
z_{\ell|k}-x_{\ell|k}
\in
\Omega
:=
\mathcal E_{\mathrm{off}}
\oplus
\mathcal E_{\mathrm{on}}
\oplus
\mathcal E_w,
\qquad
\forall\,\ell\in\mathbb N_{1:L-1}.
\]
Accordingly, we define the tightened state set
\begin{equation}\label{eq:Zl_def_TAC_final}
\mathcal Z
:=
\mathcal X\ominus\Omega.
\end{equation}
Then
\begin{equation}\label{eq:state_tightening_implication_LPV}
z_{\ell|k}\in\mathcal Z
\ \Longrightarrow \
x_{\ell|k}\in\mathcal X,
\qquad
\forall\,\ell\in\mathbb N_{1:L-1}.
\end{equation}

Similarly, the auxiliary input $u^{\mathrm a}_{\ell|k}
:=
u_{\ell|k}+K(\hat x_{\ell|k}-z_{\ell|k})$ differs from the planned input $u_{\ell|k}$ by the anchor-error feedback term. We therefore introduce the tightened input set
\begin{equation}\label{eq:Ul_def_TAC_final}
\hat{\mathcal U}
:=
\mathcal U\ominus (-K\mathcal E).
\end{equation}
Hence
\begin{equation}\label{eq:input_tightening_implication_LPV}
u_{\ell|k}\in\hat{\mathcal U}
\ \Longrightarrow\
u^{\mathrm a}_{\ell|k}
\in\mathcal U,
\qquad
\forall\,\ell\in\mathbb N_{0:L-2}.
\end{equation}

\begin{assum}\label{as:nonempty_tightened}
Throughout, the tightened state and input sets satisfy
$\mathcal Z\neq\varnothing$ and $\hat{\mathcal U}\neq\varnothing$.
\end{assum}

\subsection{Stability and Recursive Feasibility}\label{sec:rec-feas}
In this section, we establish recursive feasibility, closed-loop constraint
satisfaction, and practical ISS of the proposed controller. We consider
regulation about a common equilibrium $(u_s,x_s)$ of~\eqref{eq:lpv_ssa},
satisfying $x_s=A^\ast(p)x_s+B^\ast(p)u_s$ for all $p\in\mathbb P$.
By expressing the system and all data-driven and control-design objects in
deviation coordinates relative to $(u_s,x_s)$, this equilibrium can be placed
at the origin without loss of generality. Hence, throughout this section, we
set $(u_s,x_s)=(0,0)$. As a first step, we show that a common stabilizing gain
$K$ can be synthesized directly from the offline noisy data through a lossless
semidefinite program (SDP).
\begin{prop}\label{prop:lossless_grid_vertices}
Let $\mathcal I_\star^\theta$ and $\mathbb P$ be compact polytopes with vertex sets
 $
\operatorname{vert}\!\bigl(\mathcal I_\star^\theta\bigr)=\{\theta^{(s)}\}_{s=1}^{N_\theta}$ and
 $
\operatorname{vert}(\mathbb P)=\{p^{(r)}\}_{r=1}^{N_p}.
$
For each vertex pair $(\theta^{(s)},p^{(r)})$, let
$\bigl(A_{\theta^{(s)}}(p^{(r)}),B_{\theta^{(s)}}(p^{(r)})\bigr)$
denote the corresponding LPV matrices. Suppose there exist $P=P^\top\succ0$, $Y\in\mathbb R^{n_u\times n_x}$, and $\tau>0$ such that
\begin{equation}\label{eq:grid_vertex_LMI}
\begin{bmatrix}
P-\tau I & S^{(s,r)}\\
(S^{(s,r)})^\top & P
\end{bmatrix}\succeq 0,
\qquad
\forall\, s\in\mathbb N_{1:N_\theta},\ \forall\, r\in\mathbb N_{1:N_p},
\end{equation}
where
$S^{(s,r)}:=A_{\theta^{(s)}}(p^{(r)})P+B_{\theta^{(s)}}(p^{(r)})Y$.
Then, with $K:=YP^{-1}$,
\begin{equation}\label{eq:robust_contraction_all_theta_p}
(A+BK)P(A+BK)^\top-P+\tau I\preceq 0,
\forall\,(A,B)\in\mathcal M.
\end{equation}
Moreover,~\eqref{eq:grid_vertex_LMI} is lossless:~\eqref{eq:robust_contraction_all_theta_p} holds for all
$(\theta,p)\in\mathcal I_\star^\theta\times\mathbb P$,
or equivalently for all $(A,B)\in\mathcal M$, if and only if~\eqref{eq:grid_vertex_LMI} holds for all vertex pairs $(\theta^{(s)},p^{(r)})$.
\end{prop}

\begin{pf}
Define
\[
M(\theta,p):=
\begin{bmatrix}
P-\tau I & A_\theta(p)P+B_\theta(p)Y\\
(\cdot)^\top & P
\end{bmatrix}.
\]
Since $(A_\theta(p),B_\theta(p))$ depends affinely on $p$ for fixed $\theta$, the map $p\mapsto M(\theta,p)$ is affine. Hence
$M(\theta,p^{(r)})\succeq0$ for all $r\in\mathbb N_{1:N_p}$ implies
$M(\theta,p)\succeq0$ for all $p\in\mathbb P$ by convexity of the positive semidefinite cone.
Likewise, for fixed $p^{(r)}$, the map $\theta\mapsto M(\theta,p^{(r)})$ is affine, so
$M(\theta^{(s)},p^{(r)})\succeq0$ for all $s\in\mathbb N_{1:N_\theta}$ implies
$M(\theta,p^{(r)})\succeq0$ for all $\theta\in\mathcal I_\star^\theta$.
Combining both steps yields
$M(\theta,p)\succeq0$ for all $(\theta,p)\in\mathcal I_\star^\theta\times\mathbb P$,
or equivalently for all $(A,B)\in\mathcal M$.
Applying the Schur complement gives~\eqref{eq:robust_contraction_all_theta_p}. The losslessness claim follows by the same affine-convexity argument in reverse.
\end{pf}
\begin{rem}
\label{rem:common_terminal_controller}
The use of a common stabilizing terminal controller $K$ is mainly motivated by structural considerations. In particular, it avoids the quadratic scheduling dependence that would arise from a scheduling-dependent feedback $K(p)$ through the product $B(p)K(p)$, and thereby preserves the lossless convex characterization developed above.
In this sense, the common controller restriction is adopted primarily for
tractability rather than necessity. Allowing scheduling-dependent terminal
feedback is of independent interest and generally leads to weaker existence
conditions for stabilizing LPV terminal controllers; however, this would require
handling quadratic scheduling dependence in the closed-loop dynamics, especially
in the presence of noisy data, and therefore constitutes an interesting
direction for future research.
\end{rem}

Having synthesized a common stabilizing gain $K$, we next construct the two
set-valued ingredients used in the shifted-candidate argument. The first one is
an anchor-error RPI set $\mathcal E$ for the online mismatch recursion. In
particular, we seek $\mathcal E$ such that
\begin{equation}\label{eq:RPI_condition_TAC_refined}
\mathcal A_{\mathrm{cl}}\,\mathcal E \oplus \mathcal D \subseteq \mathcal E,
\end{equation}
where
\begin{equation}\label{eq:D_def_TAC_refined}
\mathcal D
:=
\mathcal E_{\mathrm{off}}
\oplus
\mathcal E_w
\oplus
(-\mathcal W).
\end{equation}

The second ingredient is a terminal RPI set under the terminal feedback
$u=Kz$. For $p\in\mathbb P$, define the scheduling-conditioned propagation
\begin{equation}\label{eq:Post_scheduling_operator}
\operatorname{Post}_{p}(\mathcal S)
:=
\operatorname{co}\bigl(\mathcal A_{\mathrm{cl}}(p)\mathcal S\bigr)
\oplus\mathcal E_{\mathrm{off}}.
\end{equation}
To maintain a polytopic representation uniformly over $\mathbb P$, define the
robust post operator
\begin{equation}\label{eq:Post_operator}
\operatorname{Post}(\mathcal S)
:=
\operatorname{co}\Bigl(\bigcup_{j=1}^{N_v} A_{\mathrm{cl}}^{(j)} \mathcal S\Bigr)
\oplus \mathcal E_{\mathrm{off}},
\end{equation}
so that $\operatorname{Post}_{p}(\mathcal S)\subseteq\operatorname{Post}(\mathcal S)$ for every
$p\in\mathbb P$. The terminal set is required to be invariant under this
robust propagation and compatible with the tightened constraints, as stated
next.

\begin{assum}
\label{as:terminal_RPI_set}
There exists a compact polytope $\mathcal Z_{\mathrm f}\subset\mathbb R^{n_x}$ such that
\begin{subequations}\label{eq:terminal_set_final}
\begin{align}
0\in\mathcal Z_{\mathrm f} &\subseteq \mathcal Z,
\label{eq:terminal_Minf_subset}
\\
K\mathcal Z_{\mathrm f} &\subseteq \hat{\mathcal U},
\label{eq:terminal_input_tight}
\\
\mathcal A_{\mathrm{cl}} \mathcal Z_{\mathrm f} \oplus \mathcal E_{\mathrm{off}}
&\subseteq \mathcal Z_{\mathrm f}.
\label{eq:terminal_RPI_condition}
\end{align}
\end{subequations}
\end{assum}

Under the common stabilizing gain $K$, the anchor-error set $\mathcal E$ can be
constructed offline via the data-driven set-iteration procedure
in~\cite{WangAngeli2026CDCsub}. Likewise, $\mathcal Z_{\mathrm f}$ can be
obtained by recursively propagating a compact seed set through
\(\operatorname{Post}\) until invariance and tightened admissibility are
certified.

The sets $\mathcal E$ and $\mathcal Z_{\mathrm f}$ provide the invariance
ingredients for the anchoring and terminal parts of the receding-horizon
construction. It remains to certify data representability of the predictions
needed by the recursive-feasibility argument. For
$\mathbf p\in\mathbb P^{L-1}$, let
$\mathcal P^\bullet(\mathbf p):=\mathbf p\odot I_\bullet$ and define the
admissible prediction set
\begin{equation}
\label{eq:def_auxiliary_behavior}
\begin{aligned}
\mathfrak C_L(\mathbf p)
\!:=\!\bigl\{&\chi=\operatorname{col}(u,z,u^p,z^p)\ \bigm|\\[-.2em]
&\chi\text{ satisfies the prediction-domain}\\[-.2em]
&\text{constraints of~\eqref{eq:robust_deepc} for }\mathbf p\bigr\}.
\end{aligned}
\end{equation}
The online anchoring constraint is treated separately because it depends on
$\hat x_k$. Define the full admissible prediction family as
\begin{equation}
\label{eq:def_full_admissible_family}
\mathfrak C_L^{\mathrm{adm}}
:=
\operatorname{cl}
\bigcup_{\mathbf p\in\mathbb P^{L-1}}
\mathfrak C_L(\mathbf p).
\end{equation}

For two preview vectors $\mathbf p,\mathbf p^+\in\mathbb P^{L-1}$, write
\begin{equation}
\label{eq:preview_compatibility}
\mathbf p^+\sim\mathbf p
\quad\Longleftrightarrow\quad
p_\ell^+=p_{\ell+1},\qquad
\ell\in\mathbb N_{0:L-3}.
\end{equation}
For each
$\chi=\operatorname{col}(u,z,u^p,z^p)\in
\mathfrak C_L(\mathbf p)$ and each
$\mathbf p^+\sim\mathbf p$, let
$\mathscr S_L(\chi,\mathbf p^+)$ denote the set of one-step continuations
$\chi^c=\operatorname{col}(u^c,z^c,(u^p)^c,(z^p)^c)$ satisfying
\begin{equation}
\label{eq:one_step_continuation}
\begin{aligned}
z_\ell^c&=z_{\ell+1},&
\ell&\in\mathbb N_{0:L-2},\\
u_\ell^c&=u_{\ell+1},&
\ell&\in\mathbb N_{0:L-3},\\
u_{L-2}^c&=Kz_{L-1},\\
z_{L-1}^c&\in
\operatorname{Post}_{p^+_{L-2}}(\{z_{L-1}\}),\\
(u^p)^c&=\mathcal P^{n_u}(\mathbf p^+)u^c,\\
(z^p)^c&=\mathcal P^{n_x}(\mathbf p^+)z^c_{[0,L-2]}.
\end{aligned}
\end{equation}
The admissible continuation family is
\begin{equation}
\label{eq:def_admissible_continuation_family}
\begin{aligned}
\mathfrak C_L^+
:=\operatorname{cl}\bigl\{&\chi^c\ \bigm|\
\mathbf p,\mathbf p^+\in\mathbb P^{L-1},\\
&\mathbf p^+\sim\mathbf p,\quad
\chi\in\mathfrak C_L(\mathbf p),\\
&\chi^c\in\mathscr S_L(\chi,\mathbf p^+),\quad
\chi^c\in\mathfrak C_L(\mathbf p^+)
\bigr\}.
\end{aligned}
\end{equation}
Assumption~\ref{as:terminal_RPI_set} ensures that the terminal append preserves
the tightened input, state, and terminal constraints.

The offline design fixes a compact target
$\mathcal K_L\subseteq\hat{\mathscr T}_L$ containing the admissible
continuation family and selects the factor $\bar\alpha$
in~\eqref{eq:simplex_regularizer} such that
\begin{equation}
\label{eq:auxiliary_compact_cover}
\mathfrak C_L^+
\subseteq
\mathcal K_L
\subseteq
\bar\alpha\,
\operatorname{co}\!\bigl(\operatorname{cols}(\hat H)\bigr).
\end{equation}
In view of~\eqref{eq:compact_cover_measured_dictionary}, this inclusion
certifies scaled-simplex representability of the admissible continuations and
introduces no additional online constraint.
\begin{rem}
\label{rem:continuation_coverage}
Covering $\mathfrak C_L^{\mathrm{adm}}$ is sufficient but not necessary for
recursive feasibility. The receding-horizon proof requires representability
only for the candidates in $\mathfrak C_L^+$. Thus, restricting coverage to
this continuation family can reduce the required scaled-simplex factor.
\end{rem}
We next state the main recursive feasibility and closed-loop
constraint satisfaction result.

\begin{thm}
\label{thm:rec_feas_and_consat_LPV}
Suppose Assumptions~\ref{as:scheduling_known}--\ref{as:terminal_RPI_set}
hold and the coverage condition~\eqref{eq:auxiliary_compact_cover} is
satisfied. If OCP~\eqref{eq:robust_deepc} is feasible at
$(\hat x_k,\mathbf p_k)$, then:
\begin{enumerate}
\item For the consecutive preview
$\mathbf p_{k+1}\sim\mathbf p_k$, the OCP remains feasible at time $k+1$, i.e.,
$\hat x_{k+1}\in\hat{\mathcal X}^{\mathrm{feas}}_{L}(\mathbf p_{k+1})$.

\item The implemented input
$u_k=u^\star_{0|k}+K(\hat x_k-z^\star_{0|k})$ satisfies
$u_k\in\mathcal U$, and the successor state satisfies
$x_{k+1}\in\mathcal X$.
\end{enumerate}
Consequently, if
$\hat x_0\in\hat{\mathcal X}^{\mathrm{feas}}_{L}(\mathbf p_0)$, then
\[
\hat x_k\in\hat{\mathcal X}^{\mathrm{feas}}_{L}(\mathbf p_k),\quad
u_k\in\mathcal U,\quad
x_{k+1}\in\mathcal X,\quad
\forall k\in\mathbb N_0.
\]
In particular, if
$x_0\in\mathcal X^{\mathrm{feas}}_{L}(\mathbf p_0)$, the above properties hold
for every admissible realization of the initial measurement noise.
\end{thm}

\begin{pf}
Fix $k\in\mathbb N_0$ and suppose that the OCP is feasible at
$(\hat x_k,\mathbf p_k)$. Let
$\bigl(z^\star_{[0|k,L-1|k]},\,u^\star_{[0|k,L-2|k]},\,g^\star\bigr)$
be an optimal solution of~\eqref{eq:robust_deepc}, and define its lifted vector
\[
\chi_k^\star
:=
\begin{bmatrix}
u^\star_{[0|k,L-2|k]}\\[.2em]
z^\star_{[0|k,L-1|k]}\\[.2em]
\mathcal P_k^{n_u}u^\star_{[0|k,L-2|k]}\\[.2em]
\mathcal P_k^{n_x}z^\star_{[0|k,L-2|k]}
\end{bmatrix}.
\]
The predictor and OCP constraints imply
$\chi_k^\star\in
\mathfrak C_L(\mathbf p_k)$. By
Assumption~\ref{as:scheduling_known}, consecutive preview windows satisfy
$\mathbf p_{k+1}\sim\mathbf p_k$. Choose
$z^c_{L-1|k+1}\in
\operatorname{Post}_{p_{L-2|k+1}}
(\{z^\star_{L-1|k}\})$ and define
\begin{equation}
\label{eq:shifted_candidate_horizon}
\begin{aligned}
z^c_{\ell|k+1}&:=z^\star_{\ell+1|k},&
\ell&\in\mathbb N_{0:L-2},\\
u^c_{\ell|k+1}&:=u^\star_{\ell+1|k},&
\ell&\in\mathbb N_{0:L-3},\\
u^c_{L-2|k+1}&:=Kz^\star_{L-1|k}.
\end{aligned}
\end{equation}
With
\[
\chi^c_{k+1}
:=
\begin{bmatrix}
u^c_{[0|k+1,L-2|k+1]}\\[.2em]
z^c_{[0|k+1,L-1|k+1]}\\[.2em]
\mathcal P_{k+1}^{n_u}u^c_{[0|k+1,L-2|k+1]}\\[.2em]
\mathcal P_{k+1}^{n_x}z^c_{[0|k+1,L-2|k+1]}
\end{bmatrix},
\]
the shifted state and input constraints follow from feasibility at time $k$,
while~\eqref{eq:terminal_set_final} ensures
$u^c_{L-2|k+1}\in\hat{\mathcal U}$ and
$z^c_{L-1|k+1}\in\mathcal Z_{\mathrm f}$. Together with the lifted-product
identities in $\chi^c_{k+1}$, this gives
$\chi^c_{k+1}\in\mathfrak C_L(\mathbf p_{k+1})$. Hence,
definitions~\eqref{eq:one_step_continuation}
and~\eqref{eq:def_admissible_continuation_family} give
$\chi^c_{k+1}\in\mathfrak C_L^+$. The coverage
condition~\eqref{eq:auxiliary_compact_cover} then yields
$\chi^c_{k+1}\in
\bar\alpha\operatorname{co}(\operatorname{cols}(\hat H))$, and hence there exists
$g^+\in\mathbb R^{N_{\mathrm{col}}}$ such that
\[
g^+\ge0,\qquad
\mathbf{1}^\top g^+=\bar\alpha,\qquad
\hat H g^+=\chi^c_{k+1}.
\]
The first two relations enforce the scaled-simplex constraint, while
$\hat H g^+=\chi^c_{k+1}$ yields the Hankel
predictor~\eqref{eq:lpv_hankel_predictor} at time $k+1$.

Finally, since~\eqref{eq:shifted_candidate_horizon} gives
$z^c_{0|k+1}=z^\star_{1|k}$ and $p_{0|k}=p_k$, the anchor-error recursion is
governed by~\eqref{eq:RPI_condition_TAC_refined}. Thus
$
z^c_{0|k+1}-\hat x_{k+1}\in\mathcal E.
$
Hence, the state and input components of $\chi^c_{k+1}$, together with $g^+$,
form a feasible solution of~\eqref{eq:robust_deepc} at time $k+1$.

For closed-loop constraint satisfaction, note that
$u^\star_{0|k}\in\hat{\mathcal U}$. The tightened input
condition~\eqref{eq:Ul_def_TAC_final} therefore yields
$u_k=u^\star_{0|k}+K(\hat x_k-z^\star_{0|k})\in\mathcal U$.
Similarly, $z^\star_{1|k}\in\mathcal Z$, and
by~\eqref{eq:state_tightening_implication_LPV}, one has
$x_{k+1}\in\mathcal X$.

The argument is independent of $k$; hence induction gives recursive
feasibility and closed-loop constraint satisfaction for all $k\in\mathbb N_0$
whenever
$\hat x_0\in\hat{\mathcal X}^{\mathrm{feas}}_{L}(\mathbf p_0)$. The statement
for $x_0\in\mathcal X^{\mathrm{feas}}_{L}(\mathbf p_0)$ follows
from~\eqref{eq:feasible_sets_def}.
\end{pf}

We now establish practical ISS of the actual LPV closed-loop on the preview-parameterized feasible domain. Throughout, the implemented input is
$u_k=u^\star_{0|k}+K(\hat x_k-z^\star_{0|k})$. The optimal value function is used as Lyapunov
candidate. The terminal penalty $P_L\succ0$ is chosen to be parameter independent,
consistently with the stabilizing gain $K$, as a common quadratic Lyapunov
matrix for the closed-loop family so that
\begin{equation}\label{eq:primal_lyap_final_LPV}
A_{\mathrm{cl}}^\top P_L A_{\mathrm{cl}}-P_L+S\prec0,
\qquad
\forall\,A_{\mathrm{cl}}\in\mathcal A_{\mathrm{cl}},
\end{equation}
where $S:=Q+K^\top R K$. Thus, $P_L$ provides a uniform terminal-cost decrease
for all $A_{\mathrm{cl}}\in\mathcal A_{\mathrm{cl}}$. A convenient choice is to
take $P_L$ as a suitably scaled inverse of the SDP Lyapunov certificate $P$,
thereby converting the dual decay condition certified by the data-driven design
into the primal terminal Lyapunov inequality required
here~\cite{WangAngeli2026TRDDPC}. With this choice in place, consider the
optimal finite-horizon value function
\begin{equation}\label{eq:Vz_def_final}
V_L^\star(\hat x_k,\mathbf p_k)
:=
\sum_{\ell=0}^{L-2}
J_{\mathrm s}\!\bigl(z^\star_{\ell|k},u^\star_{\ell|k}\bigr)
+
J_{\mathrm f}\!\bigl(z^\star_{L-1|k}\bigr).
\end{equation}
To derive a Lyapunov estimate uniformly over scheduling previews, we impose the
following standard quadratic cost controllability condition.

\begin{assum}
\label{as:uniform_cost_controllability}
There exists a constant $c_V>0$, independent of the scheduling preview, such
that
\begin{equation}
\label{eq:uniform_value_upper_bound}
V_L^\star(\hat x,\mathbf p)
\le
c_V\|\hat x\|^2,
\qquad
\forall\,\mathbf p\in\mathbb P^{L-1},\quad
\forall\,\hat x\in
\hat{\mathcal X}^{\mathrm{feas}}_L(\mathbf p).
\end{equation}
\end{assum}
For each fixed preview, OCP~\eqref{eq:robust_deepc} is a convex parametric
QP in $\hat x$;
under standard regularity conditions, its value function is continuous and
piecewise quadratic~\cite{RawlingsMayneDiehl2017}. Uniformity is required
because the preview parameterizes the feasible Hankel trajectories, even though
the objective does not depend on it.

The next theorem shows that $V_L^\star$ is a practical ISS Lyapunov function for the actual closed-loop.
\begin{thm}\label{thm:ISS_value}
Suppose Assumptions~\ref{as:scheduling_known}--\ref{as:uniform_cost_controllability}
hold, the coverage
condition~\eqref{eq:auxiliary_compact_cover} is satisfied, and $P_L$
satisfies~\eqref{eq:primal_lyap_final_LPV}. Set
$\epsilon:=\max_{w\in\mathcal W}\|w\|$. Then there exist
class-$\mathcal K_\infty$ functions $\alpha_1,\alpha_2$, a constant
$\kappa>0$, and class-$\mathcal K$ functions $\sigma,\sigma_\Delta$ such
that, for any scheduling sequence $p:\mathbb N_0\to\mathbb P$ satisfying
Assumption~\ref{as:scheduling_known}, any
$x_0\in\mathcal X^{\mathrm{feas}}_{L}(\mathbf p_0)$ and all
$k\in\mathbb N_0$,
\begin{align}
\alpha_1(\|x_k\|)-\sigma(\epsilon)
\le
V_L^\star(\hat x_k,\mathbf p_k)
\le
\alpha_2(\|x_k\|)+\sigma(\epsilon),
\label{eq:ISS_thm_bounds_value}
\\
V_L^\star(\hat x_{k+1},\mathbf p_{k+1})
-V_L^\star(\hat x_k,\mathbf p_k)
\le
-\,\kappa\,\|x_k\|^2
+\sigma_\Delta(\epsilon).
\label{eq:ISS_thm_diff_value}
\end{align}
Consequently, there exist $\beta\in\mathcal{KL}$ and $\Gamma\in\mathcal K$ such that
\begin{equation}\label{eq:ISS_state_bound_value}
\|x_k\|
\le
\beta(\|x_0\|,k)+\Gamma(\epsilon),
\qquad
\forall\,k\in\mathbb N_0.
\end{equation}
Hence, the true LPV closed-loop is practically input-to-state stable on its
recursively feasible domain with respect to the common offline and online
measurement-noise level $\epsilon$.
\end{thm}

\begin{pf}
By Theorem~\ref{thm:rec_feas_and_consat_LPV}, the OCP remains feasible along any closed-loop trajectory starting from
$\mathcal X^{\mathrm{feas}}_{L}(\mathbf p_0)$. Under the terminal law $u=Kz$,
the appended terminal state of the shifted candidate satisfies
\[
z^+=A_{\mathrm{cl}}z+\delta,
\quad
A_{\mathrm{cl}}\in
\mathcal A_{\mathrm{cl}}(p_{L-2|k+1})
\subseteq\mathcal A_{\mathrm{cl}},
\quad
\delta\in\mathcal E_{\mathrm{off}}.
\]
Using~\eqref{eq:primal_lyap_final_LPV} and Young's inequality, one obtains a
computable class-$\mathcal K$ function $c_{\mathrm f}$ such that
\[
J_{\mathrm f}(z^+)-J_{\mathrm f}(z)
\le
-\,z^\top S z+c_{\mathrm f}(\epsilon),
\qquad
\forall\,z\in\mathcal Z_{\mathrm f}.
\]
Combining this estimate with the shifted candidate from the proof of
Theorem~\ref{thm:rec_feas_and_consat_LPV} gives
\[
V_L^\star(\hat x_{k+1},\mathbf p_{k+1})
-V_L^\star(\hat x_k,\mathbf p_k)
\le
-\,J_{\mathrm s}\!\bigl(z^\star_{0|k},u^\star_{0|k}\bigr)+c_{\mathrm f}(\epsilon).
\]
Since $Q\succ0$ and $R\succ0$, there exists $\underline q>0$ such that
$J_{\mathrm s}(z,u)\ge\underline q\|z\|^2$. Hence,
\begin{equation}\label{eq:Vz_shift_ineq_coercive_TAC}
V_L^\star(\hat x_{k+1},\mathbf p_{k+1})
-V_L^\star(\hat x_k,\mathbf p_k)
\le
-\,\underline q\,\|z^\star_{0|k}\|^2+c_{\mathrm f}(\epsilon).
\end{equation}

Moreover, the anchoring constraint together with $\hat x_k=x_k+w_k$ and
$w_k\in\mathcal W$ implies, for some $\beta_E>0$,
$\|z^\star_{0|k}-x_k\|\le \beta_E\epsilon$. Consequently,
$\|z^\star_{0|k}\|^2\ge
\frac{1}{2}\|x_k\|^2-\beta_E^2\epsilon^2$.
The nonnegativity of the costs, Assumption~\ref{as:uniform_cost_controllability},
and $\|\hat x_k\|^2\le2\|x_k\|^2+2\epsilon^2$ then give
\[
\frac{\underline q}{2}\|x_k\|^2-\underline q\beta_E^2\epsilon^2
\le
V_L^\star(\hat x_k,\mathbf p_k)
\le
2c_V\|x_k\|^2+2c_V\epsilon^2.
\]
Thus,~\eqref{eq:ISS_thm_bounds_value} holds with
$\alpha_1(s):=\frac{\underline q}{2}s^2$,
$\alpha_2(s):=2c_Vs^2$, and, for example,
$\sigma(s):=(\underline q\beta_E^2+2c_V)s^2$.
Combining the anchor estimate with~\eqref{eq:Vz_shift_ineq_coercive_TAC}
gives~\eqref{eq:ISS_thm_diff_value} with
$\kappa:=\underline q/2$ and
$\sigma_\Delta(s):=\underline q\beta_E^2s^2+c_{\mathrm f}(s)$.
The bound~\eqref{eq:ISS_state_bound_value} then follows from standard
discrete-time ISS Lyapunov arguments~\cite{JiangWang2001Automatica}.
\end{pf}

\section{Example}\label{sec:examples}
We consider a unit-mass inverted oscillator with externally scheduled
stiffness. Forward-Euler discretization gives
\begin{equation}\label{eq:lpv_ss_external_oscillator}
\begin{aligned}
x_{k+1}&=A(p_k)x_k+Bu_k,\qquad \hat x_k=x_k+w_k,\\
A(p)&=
\begin{bmatrix}
1&T_s\\
 T_s(k_0+k_1p)&1-T_sc_d
\end{bmatrix},\qquad
B=\begin{bmatrix}0\\T_s\end{bmatrix},
\end{aligned}
\end{equation}
where $x_1$ and $x_2$ denote displacement and velocity, respectively. We set
$T_s=0.05$, $k_0=1$, $k_1=0.35$, and $c_d=0.8$. The prediction length is
$L=10$. The normalized stiffness command $p_k$ is available over the prediction
horizon. For this numerical experiment, we choose the following periodic
command:
\begin{equation}\label{eq:external_periodic_scheduling}
(p_0,\ldots,p_4)
\!=\!
(0.0927,\!0.8000,\!0.4017,\!-0.3444,\!-0.9500).
\end{equation}
Thus $p_k\in\mathbb P:=[-1,1]$, and the preview windows at consecutive
sampling instants satisfy~\eqref{eq:preview_compatibility}. The joint spectral
radius of $\mathcal A:=\{A(p)\mid p\in\mathbb P\}$ is
$\rho_{\mathrm{jsr}}(\mathcal A)\approx1.041441>1$. Moreover,
$\rho(A(p_4)\cdots A(p_0))\approx1.181108>1$ for the command
in~\eqref{eq:external_periodic_scheduling}; hence, the system is open-loop
unstable under this scheduling sequence. Periodicity is specific to this
numerical setup and is not required by the proposed controller.

The state and input constraints and the measurement noise set are
\begin{equation}\label{eq:external_example_sets}
\begin{aligned}
\mathcal X&:=
\{x\in\mathbb R^2\mid |x_1|\le1,\ |x_2|\le2\},\\
\mathcal U&:=\{u\in\mathbb R\mid |u|\le4\},\\
\mathcal W&:=
\{w\in\mathbb R^2\mid |w_1|\le10^{-3},\ |w_2|\le10^{-3}\}.
\end{aligned}
\end{equation}
The initial condition is $x_0=[-0.7\ \ 0.5]^\top$, and the control objective is
to stabilize the origin. We use $Q=\operatorname{diag}(6,1)$ and $R=0.05$.

The offline input-scheduling-state data were arranged as length-$L$ records,
with the initial phase of~\eqref{eq:external_periodic_scheduling} varied across
the data records. Collecting offline trajectories over a region larger than
the online constraint set can reduce the required coverage factor
$\bar\alpha$, provided that the data-collection experiment remains safely
implementable. Data collection was permitted within $|x_1^d|\le4$,
$|x_2^d|\le8$, and $|u^d|\le16$, while the OCP continued to
enforce~\eqref{eq:external_example_sets}. To further reduce the finite-data covering
factor, the numerical design employs the admissible one-step continuation
family generated by the residual-restricted OCP prediction domain, rather than
the full admissible prediction family. A total of 400 measured trajectory
columns were retained after append-only data enrichment. Their negatives were
also included, giving $N_{\mathrm{col}}=800$ and
ensuring~\eqref{eq:origin_relative_interior_zhat} relative to the data span.

\begin{figure}[tbp]
    \centering
    \includegraphics[width=\columnwidth]{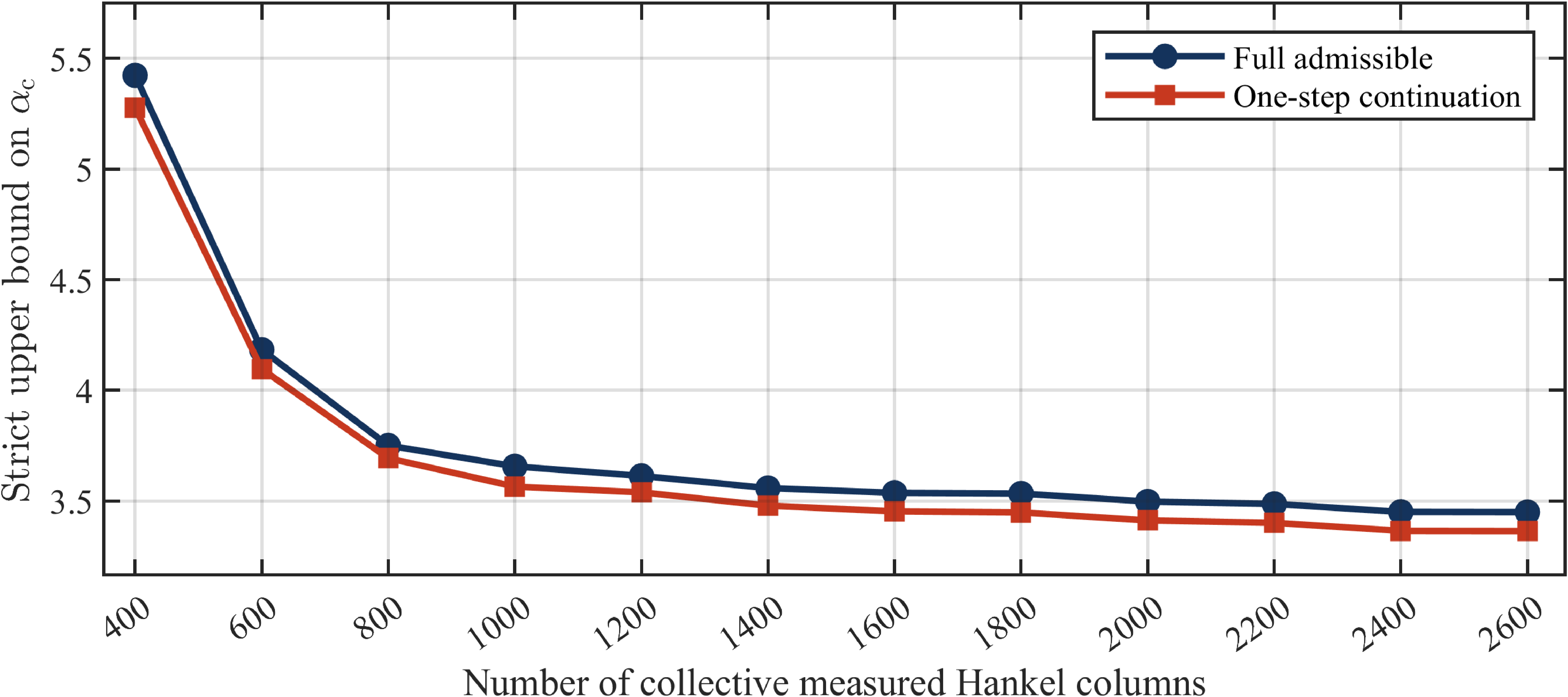}
    \caption{Strict covering-factor upper bounds for the full admissible
    prediction family and the admissible one-step continuation family.}
    \label{fig:coverage_comparison}
\end{figure}

Using the same measured dictionary and residual restriction, the full
admissible prediction family and the admissible one-step continuation family
give the certified strict upper bounds
$3.751024153$ and $3.694011714$, respectively. The latter is used online, i.e.,
$\bar\alpha=3.694011714$. Fig.~\ref{fig:coverage_comparison} shows that both
bounds decrease as trajectory columns are appended, while the continuation
target consistently requires the smaller factor. This illustrates a
coverage--computation tradeoff. A smaller $\bar\alpha$ contracts the offline
mismatch set $\mathcal E_{\mathrm{off}}=\bar\alpha\mathcal D_A$ and can thereby
reduce constraint tightening, enlarge the admissible terminal region, and
provide additional margin against measurement noise. Conversely,
$N_{\mathrm{col}}$ is the dimension of the online Hankel coefficient vector
$g$, so a larger dictionary increases the online QP burden. Beyond
$N_{\mathrm{col}}=800$, the improvement becomes comparatively modest:
increasing the dictionary to 2600 columns increases the dimension of $g$ by
approximately $225\%$, while reducing the continuation upper bound from
$3.694011714$ to $3.364901515$, i.e., by approximately $9\%$. We therefore use
$N_{\mathrm{col}}=800$ in the closed-loop study as a compromise between
tightening reduction and online problem size.

\begin{figure}[tbp]
    \centering
    \subfloat[]{%
        \includegraphics[width=0.48\columnwidth]{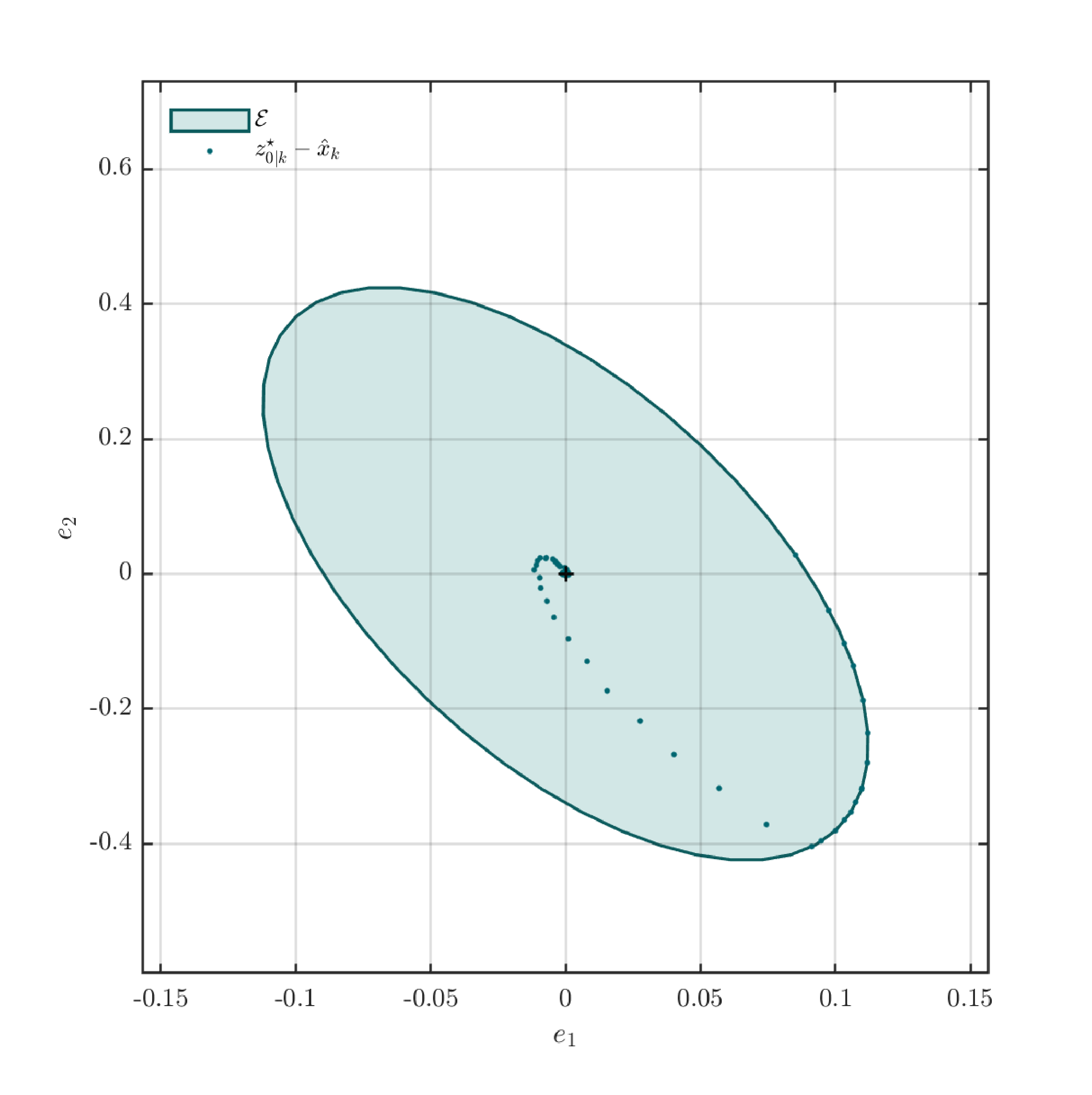}%
        \label{fig:initial_rpi_set}%
    }\hfil
    \subfloat[]{%
        \includegraphics[width=0.48\columnwidth]{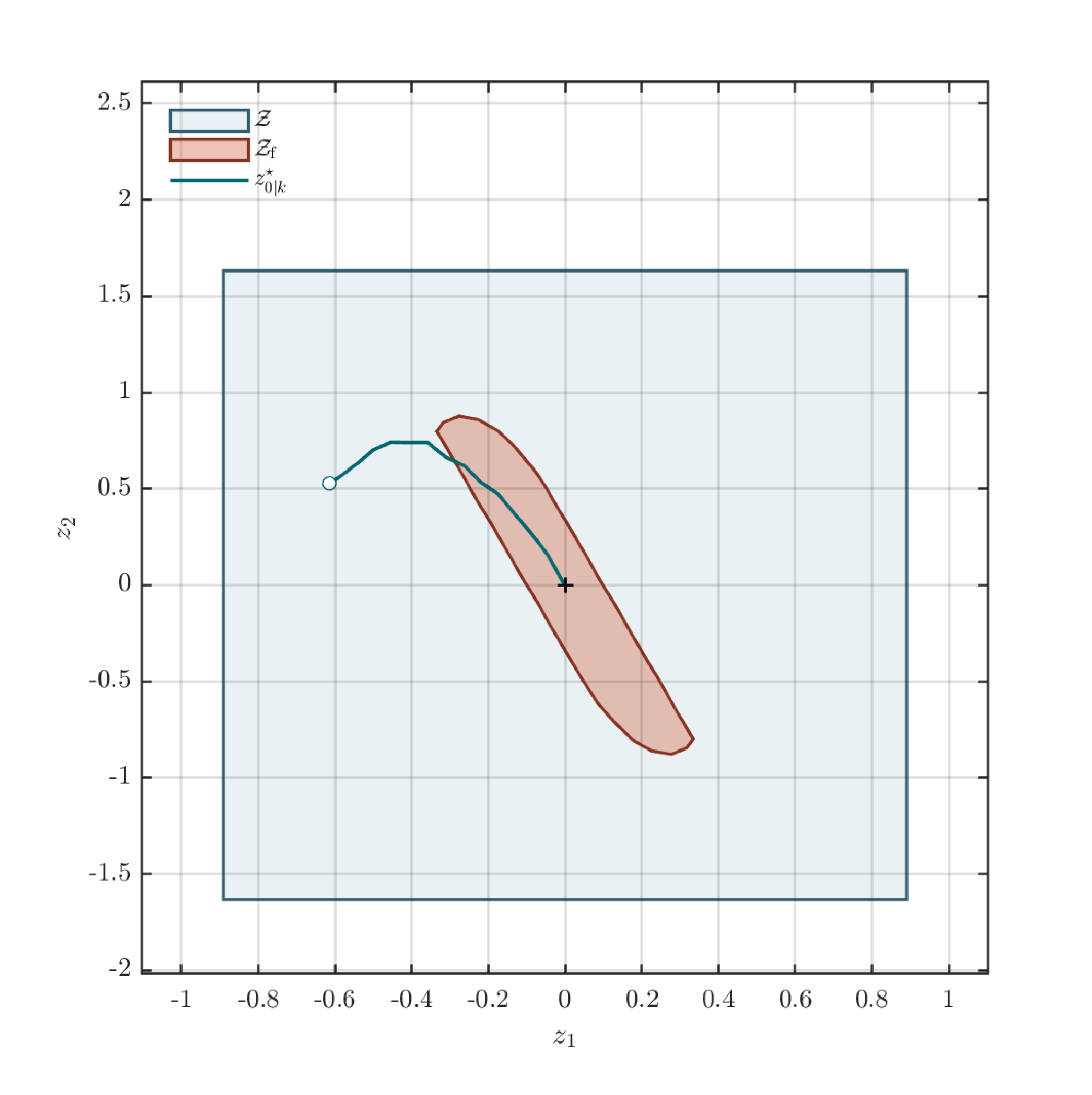}%
        \label{fig:tightened_terminal_sets}%
    }
    \caption{Offline-computed RPI and tightened sets. Fig.~\protect\subref{fig:initial_rpi_set} shows $\mathcal E$ and the realized anchor errors $z_{0|k}^\star-\hat x_k$. Fig.~\protect\subref{fig:tightened_terminal_sets} shows $\mathcal Z$, $\mathcal Z_{\mathrm f}$, and the optimal initial predicted states $z_{0|k}^\star$.}
    \label{fig:sets_geometry}
\end{figure}

Using the same noisy data, we constructed the refined model-consistency
polytope $\mathcal I_\star^\theta$ and its induced model family $\mathcal M$.
The common gain
$K=\begin{bmatrix}-19.4060&-5.7020\end{bmatrix}$ was certified over this family
through~\eqref{eq:robust_contraction_all_theta_p}. The corresponding mismatch
bounds were then used to compute an anchor-error RPI polytope $\mathcal E$
satisfying~\eqref{eq:RPI_condition_TAC_refined}, followed by the tightened
sets $\mathcal Z$ and $\hat{\mathcal U}$
in~\eqref{eq:Zl_def_TAC_final} and~\eqref{eq:Ul_def_TAC_final}. Finally,
$\mathcal Z_{\mathrm f}$ was computed to satisfy the terminal
conditions~\eqref{eq:terminal_set_final}, while
$P_L=\begin{bmatrix}108.5649&17.1673\\17.1673&7.5650\end{bmatrix}$
was chosen to satisfy the common Lyapunov
inequality~\eqref{eq:primal_lyap_final_LPV}. The data-derived
sets $\mathcal E$, $\mathcal Z$, and $\mathcal Z_{\mathrm f}$ are illustrated
in Fig.~\ref{fig:sets_geometry}. In practical implementations of the online
OCP, the scaled simplex constraint~\eqref{eq:simplex_regularizer} can be relaxed
to
\begin{equation}\label{eq:simplex_regularizer_relaxed}
g\in\Delta_{r}^{N_{\mathrm{col}}},
\qquad
\Delta_{r}^{N_{\mathrm{col}}}
:=
\{\,g\in\mathbb{R}^{N_{\mathrm{col}}}\mid g\ge 0,\ \mathbf{1}^\top g\le \bar\alpha\,\}.
\end{equation}
This relaxation enlarges the admissible coefficient set and therefore cannot
reduce feasibility. In particular, all recursive feasibility and closed-loop
constraint satisfaction results established above remain valid, since the
corresponding arguments rely only on the existence of a feasible coefficient
vector under the original equality-constrained simplex, and the relaxed
set~\eqref{eq:simplex_regularizer_relaxed} contains that set as a subset.
Moreover, penalizing the coefficient vector $g$ can mitigate the effect of
noisy Hankel data on open-loop predictions~\cite{9705109}. In the numerical
implementation, we augment the OCP objective with the quadratic regularization
term $\lambda_g\|g\|_2^2$, where $\lambda_g=10^{-2}$.

\begin{figure}[tbp]
    \centering
    \subfloat[]{%
        \includegraphics[width=\columnwidth]{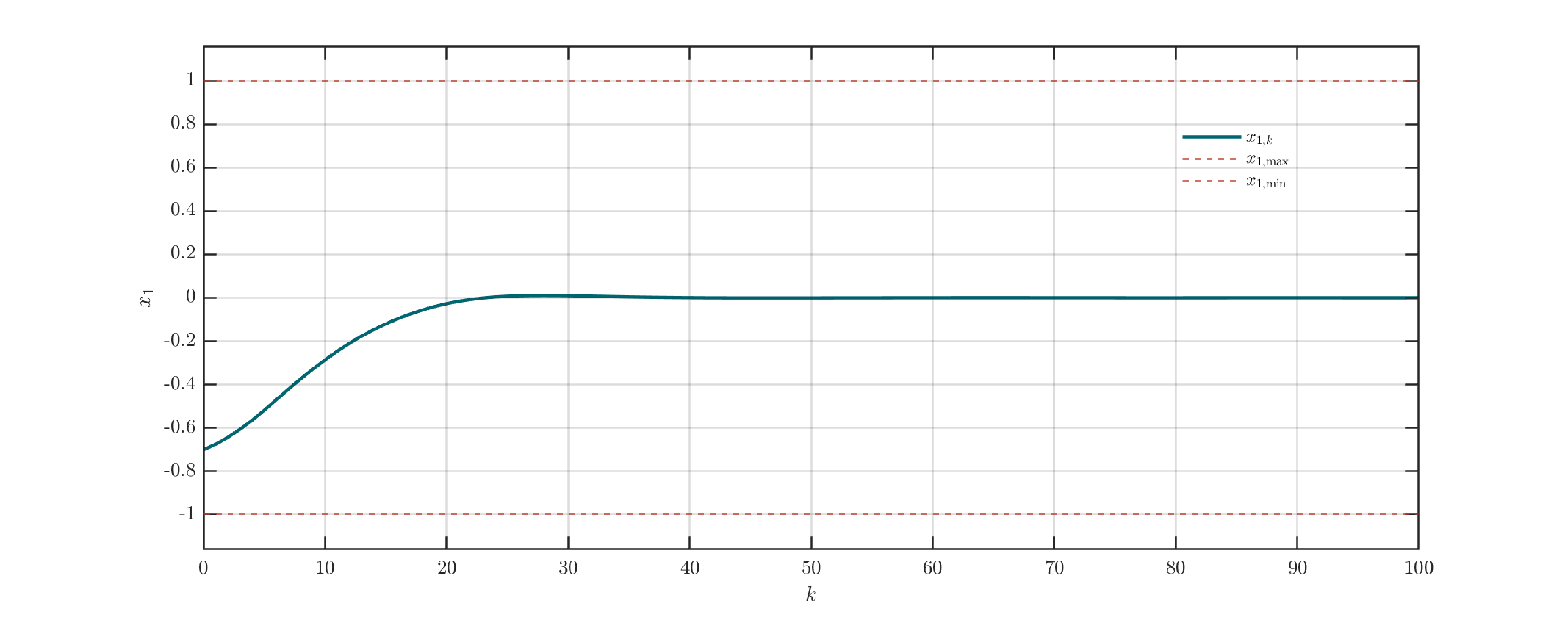}%
        \label{fig:cl_x1}%
    }\\[0.5ex]%
    \subfloat[]{%
        \includegraphics[width=\columnwidth]{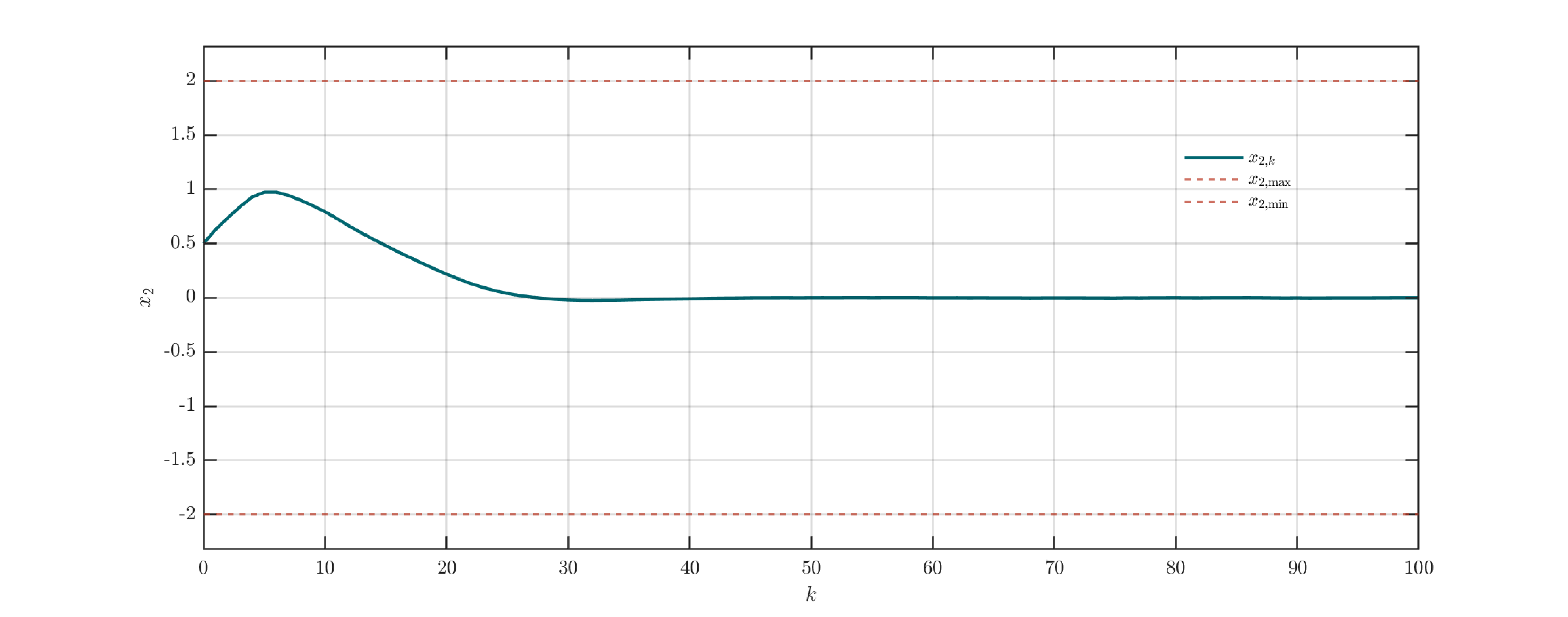}%
        \label{fig:cl_x2}%
    }\\[0.5ex]%
    \subfloat[]{%
        \includegraphics[width=\columnwidth]{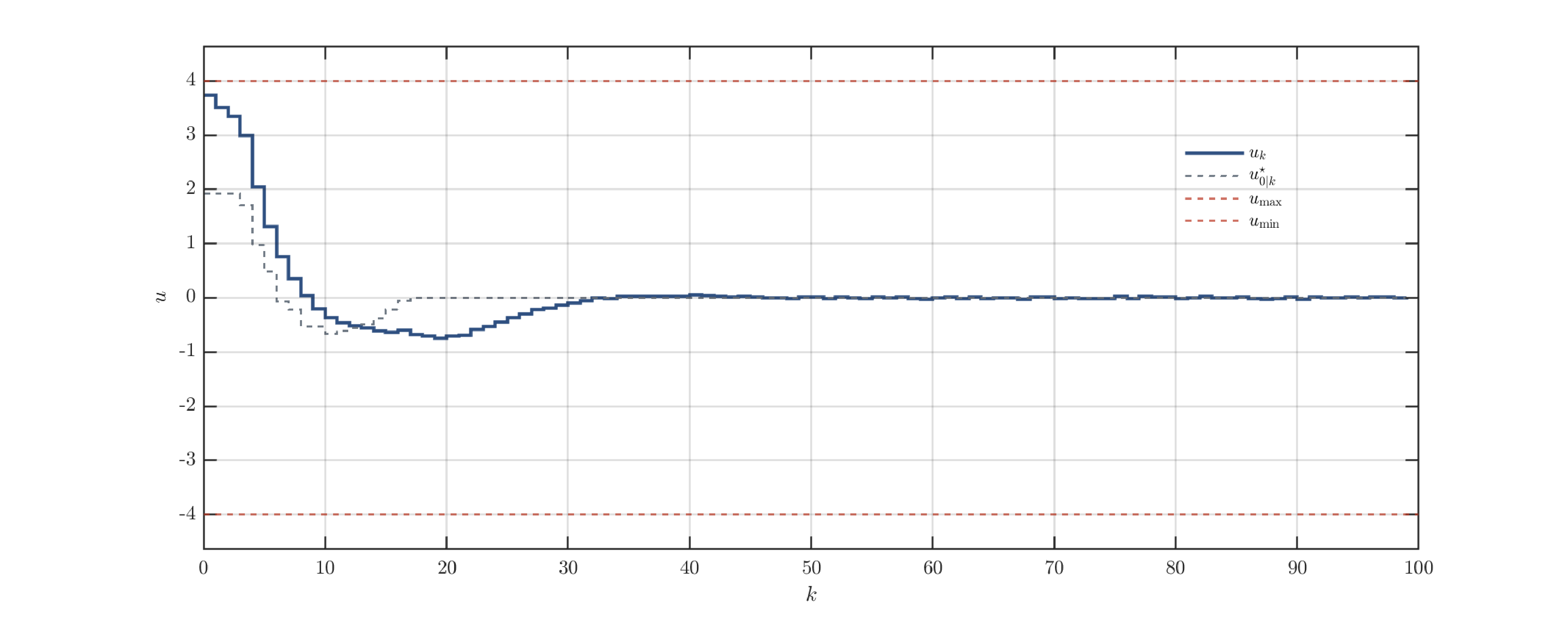}%
        \label{fig:cl_input}%
    }
    \caption{Closed-loop trajectories of the externally scheduled inverted oscillator.
    Figs.~\protect\subref{fig:cl_x1} and~\protect\subref{fig:cl_x2} show the
    true states together with the original constraints;
    Fig.~\protect\subref{fig:cl_input} shows the applied and planned inputs.}
    \label{fig:cl_trajectories}
\end{figure}

The closed-loop state and input trajectories over 100 sampling instants are
shown in Fig.~\ref{fig:cl_trajectories}. The anchor errors remained in
$\mathcal E$, and the tube correction
$K(\hat x_k-z_{0|k}^\star)$ accounts for the difference between the planned
and applied inputs in Fig.~\ref{fig:cl_input}. The applied input and true state
respected the original sets $\mathcal U$ and $\mathcal X$, respectively, as
ensured by the tightened constraints. After the transient, the state remains
in a small noise-dependent neighborhood of the origin, while the planned and
applied inputs settle near the equilibrium input $u_s=0$. Thus, the closed
loop is stabilized around the prescribed equilibrium despite measurement
noise, consistently with the practical ISS conclusion of
Theorem~\ref{thm:ISS_value}.

\section{Conclusion}\label{sec:conclusions}
This paper developed an LPV-TRDDPC scheme
for unknown LPV systems using offline noisy input-scheduling-state data. The
scheme handles prediction mismatch through tube
tightening and uses a coverage condition to retain scaled-simplex Hankel
representability. The resulting tightening ingredients were computed directly
from noisy data. The online
implementation reduces to a receding-horizon QP. Recursive
feasibility, closed-loop constraint satisfaction,
and practical ISS were established. The externally scheduled inverted-oscillator
example illustrated closed-loop stabilization under measurement
noise and satisfaction of the imposed constraints.

A distinctive feature of the proposed framework is that it does not aim to preserve a representation of the entire unconstrained behavior of the unknown LPV system. Instead, it focuses on the constrained behavior relevant to the control task. This constrained-behavior viewpoint is compatible with the LTI case and with LPV embeddings of certain nonlinear dynamics, providing a structured basis for robust direct data-driven predictive control beyond the LTI setting.

Several directions deserve further investigation. In particular, although the
proposed framework provides robustness guarantees under bounded noise, it still
involves conservatism. Reducing this conservatism remains an important topic
for future work. It is also of interest to further extend the proposed framework so as to broaden the applicability of robust data-driven predictive control to more general nonlinear systems.

\bibliographystyle{plain}
\bibliography{autosam}

\appendix
\section{Offline consistency refinement}
\label{app:gamma-tight-LPV}

Starting from the enclosure $\mathcal I_0^\theta$
in~\eqref{eq:initial_model_enclosure}, define
\begin{equation}\label{eq:rho_i_def_app_LPV}
\gamma_{i,\ell}^{(r)}
:=
\max_{\theta\in\mathcal I_r^\theta}
\|A_\theta(p^{d,i}_{\ell})\|_{\mathcal W}.
\end{equation}
Stack these values in $\boldsymbol\gamma^{(r)}$ and set
\begin{equation}\label{eq:nested_consistency_refinement}
\mathcal I_{r+1}^\theta
:=
\mathcal I_r^\theta
\cap
\mathcal I^\theta(\boldsymbol\gamma^{(r)}),
\qquad r\in\mathbb N_0.
\end{equation}
By the properties of $\mathcal W$ stated after~\eqref{eq:disturbance_polytope}
and compactness of $\mathcal I_0^\theta$, the maxima
in~\eqref{eq:rho_i_def_app_LPV} are finite.

An induction shows that, for every $r\in\mathbb N_0$,
\begin{equation}\label{eq:nested_refinement_properties}
\theta^\ast\in\mathcal I_{r+1}^\theta
\subseteq\mathcal I_r^\theta,
\qquad
\boldsymbol\gamma^{(r+1)}\le\boldsymbol\gamma^{(r)},
\end{equation}
and each $\mathcal I_r^\theta$ is a nonempty compact polytope. Indeed,
Proposition~\ref{prop:initial_model_enclosure} gives
$\theta^\ast\in\mathcal I_0^\theta$. Suppose
$\theta^\ast\in\mathcal I_r^\theta$.
Then~\eqref{eq:rho_i_def_app_LPV} gives
$A^\ast(p_\ell^{d,i})\mathcal W
\subseteq\gamma_{i,\ell}^{(r)}\mathcal W$
for every $(i,\ell)\in\mathscr J$. Hence the residual associated with the true
parameter belongs to $\mathcal D(\gamma_{i,\ell}^{(r)})$, and therefore
$\theta^\ast\in\mathcal I^\theta(\boldsymbol\gamma^{(r)})$. It follows
from~\eqref{eq:nested_consistency_refinement} that
$\theta^\ast\in\mathcal I_{r+1}^\theta\subseteq\mathcal I_r^\theta$.
Compactness and polyhedrality are preserved under intersection, while
maximization over the nested sets gives
$\boldsymbol\gamma^{(r+1)}\le\boldsymbol\gamma^{(r)}$.

Fix any finite $r_\star\in\mathbb N_0$ and define
\begin{equation}\label{eq:refined_model_set_def}
\boldsymbol\gamma^\star:=\boldsymbol\gamma^{(r_\star)},
\qquad
\mathcal I_\star^\theta:=\mathcal I_{r_\star+1}^\theta.
\end{equation}
The properties in~\eqref{eq:nested_refinement_properties}
yield~\eqref{eq:refined_model_set_properties}. Moreover,
\begin{equation}\label{eq:refined_model_set_gain}
A_\theta(p_\ell^{d,i})\mathcal W
\subseteq
\gamma_{i,\ell}^\star\mathcal W,
\quad
\forall\theta\in\mathcal I_\star^\theta,
\quad
\forall(i,\ell)\in\mathscr J.
\end{equation}
Consequently, any finite $r_\star$ gives a valid model set and inflation
vector. When $\mathcal W$ is polytopic, each term
$\|A_\theta(p^{d,i}_{\ell})\|_{\mathcal W}$ is computable via finitely many linear
programs, so the refinement can be carried out offline.

After $\boldsymbol\gamma^\star$ has been fixed, the image set $\mathcal W_A$
used in~\eqref{eq:WA_star_uniform_LPV} can be computed independently of the
refinement. For instance, given candidate facet normals
$\{f_r\}_{r\in\mathbb N_{1:N_A}}$, one may choose the polytopic outer approximation
\[
\mathcal W_A
:=
\{y\in\mathbb R^{n_x}: f_r^\top y\le b_{A,r},
r\in\mathbb N_{1:N_A}\},
\]
where $h_{\mathcal W}(v):=\max_{w\in\mathcal W}v^\top w$ denotes the support
function of $\mathcal W$, and
\[
b_{A,r}
\ge
\max_{\substack{\theta\in\mathcal I_\star^\theta\\
p\in\mathbb P}}
h_{\mathcal W}\!\bigl(A_\theta(p)^\top f_r\bigr),
\qquad r\in\mathbb N_{1:N_A}.
\]
This support function construction includes componentwise and
direction-dependent bounds as special cases. The scalar dilation
$\mathcal W_A=\gamma^\star\mathcal W$ is recovered by taking
$\gamma^\star$ no smaller than the corresponding worst-case
$\mathcal W$-gain; if $\mathcal W$ is centrally symmetric, then
$\mathcal W\oplus(-\mathcal W_A)=(1+\gamma^\star)\mathcal W$.

\end{document}